\documentclass[proof]{WileyASNA-v1}

\usepackage[utf8]{inputenc}
\usepackage{calc}
\usepackage{anyfontsize}
\usepackage{hyperref}
\usepackage{endnotes}

\newlength{\okinalen}
\newcommand{\okina}{\hbox to.666\okinalen{\hss`\hss}}

\articletype{Article Type}

\received{2025}
\revised{2025}
\accepted{2025}

\begin{document}

\title{Radial Velocity Monitoring and Analysis of Gaia Astrometry of Selected Intermediate Mass Stars to Constrain Their Multiplicity Status}

\author[1]{J. B\"{a}tz*}
\author[1]{M. Mugrauer}
\author[1]{K.-U. Michel}
\author[1]{J. Reichert}
\author[1]{A. Tschirschky}
\author[1]{L. Pietsch}
\author[1]{F. Edelmann}
\author[1]{R. Neuh\"{a}user}

\authormark{J. B\"{a}tz \textsc{et al}}

\address[]{\orgdiv{Astrophysical Institute and University Observatory}, \orgname{Friedrich-Schiller-University Jena}, \orgaddress{\state{Schillerg\"{a}{\ss}chen 2, 07745 Jena}, \country{Germany}}}

\corres{*Janine B\"{a}tz. \email{baetz.janine@gmail.com}}

\abstract{We present new radial velocity measurements of 13 selected intermediate mass stars (2 -- 6\,M$_\odot$). The measurements were performed between 29 April and 6 September 2024 at the University Observatory Jena using the \'echelle spectrograph FLECHAS. The radial velocity of eight stars was found to be constant during our spectroscopic monitoring, namely: 17\,Dra\,A, HD\,148374, HD\,169487\,A, 57\,Cnc, $\gamma$\,And, HD\,11031, $\kappa$\,And, and $\lambda$\,Cas. In contrast, the radial velocity of five stars showed significant variability throughout or spectroscopic observation, namely: 7\,CrB\,A, 7\,CrB\,B, HD\,214007, $\iota$\,Her, and HD\,201433\,A. In all these cases, Keplerian orbital solutions were fitted to the observational data and the orbital elements of these spectroscopic binary systems were determined. In addition, we searched for wide companions of our targets using the third data release from ESA's Gaia mission, in order to determine the multiplicity status of these stars and contribute to the census of bright, nearby multiple stars.}

\keywords{techniques: radial velocities, methods: observational, binaries: general, binaries: spectroscopic, stars: individual (17\,Dra\,A, HD\,148374, HD\,169487\,A, 57\,Cnc, $\gamma$\,And, HD\,11031, $\kappa$\,And, $\lambda$\,Cas, 7\,CrB\,A, 7\,CrB\,B, HD\,214007, $\iota$\,Her, HD\,201433\,A)}

\maketitle

\section{Introduction}\label{sec:intro}

In \cite{baetz2025} about 600000 stars with absolute magnitudes \mbox{$M_{\text{G}} \leq 0.65$\,mag}, located within a radius of 1.7\,kpc around the Sun, were retrieved from the third data release of the ESA Gaia mission \citep[Gaia DR3 from here on, see][]{gaiadr3}, to detect OB-type runaway stars\footnote{Unusually fast-moving early-type stars, which have attained their high space velocities either by dynamical ejection or by supernova explosion in multiple star systems.} originating from the Scorpius-Centaurus-Lupus association among them. This sample contains a large number of stars with no known radial velocity (RV) in the Gaia DR3, the Extended Hipparcos Compilation \citep{anderson2012} or the Simbad database \citep{wenger2000}, which are however needed to trace their trajectories back in time. In order to obtain at least some additional RV measurements, a subset of these stars was therefore selected for observation at the University Observatory Jena. Since most of the runaway stars are slower than the peak velocities of runaway stars of about 200\,km/s \citep{blaauw1961}, the most promising runaway star candidates found in \citep{baetz2025} were expected to still be located relatively close to the Scorpius-Centaurus-Lupus association. Therefore, only stars within a defined radius of approximately 166\,pc around the Sun were considered for observation. This resulted in a subsample of 105 stars, 13 of which can be observed at the University Observatory Jena. The properties of these targets are summarized in \mbox{Table\,\ref{tab:obs}\hspace{-1.5mm}}. It should be noted that the stars presented in this paper were then excluded as runaway stars from the Scorpius-Centaurus-Lupus association according to \cite{baetz2025} -- partly because of the RV measured here. All targets were observed with the \'echelle spectrograph FLECHAS \citep[][Fibre Linked ECHelle Astronomical Spectrograph] {mugrauer2014} operated at the 0.9\,m telescope of the University Observatory Jena to determine their RV. For stars whose RV remained constant during the observations, the mean and standard deviation of their RV were calculated and are presented in this paper. For stars with a significantly variable RV, we attempted to fit a best-fitting Keplerian orbital solution to the RV data to determine the orbital elements of these systems, including their systemic velocity.

\begin{table*}[h!]  \caption{The observed targets together with their right ascension $\alpha$, declination $\delta$, apparent G-band brightness, Gaia FLAME mass estimate $mass$, effective temperature $T_\text{eff}$ and surface gravity $\log(g)$, as listed in the Gaia DR3. The geometric distance $dist$ of all targets was taken from \cite{bailerjones2021}. In addition, the detector integration time (DIT) and the number of observation epochs $N_\text{Obs}$ are given for each target. The range of $T_\text{eff}$ of all targets corresponds to spectral types B5 -- K0.}

\centering
\resizebox{\hsize}{!}{
\begin{tabular}{lccccccccc}
\toprule
Target  & $\alpha$           & $\delta$           & $G$       & $dist$   & $mass$           &  $T_\text{eff}$  &  $\log(g[\text{cm}/\text{s}^2])$  & DIT   & $N_\text{Obs}$\\
& $[$hh:mm:ss.ss$]$ & $[$dd:mm:ss.s$]$ & $[$mag$]$ & $[$pc$]$ & $[$M$_\odot$$]$  &  $[$K$]$         &                                   & $[$s$]$ & \\
\midrule
17\,Dra\,A &     16 36 13.72 & +52 55 27.9 & $5.402 \pm 0.003$ & $126.79^{+1.44}_{-1.82}$   & $3.09^{+0.05}_{-0.04}$ & $10832^{+56}_{-51}$ & $3.76^{+0.01}_{-0.01}$ & 150 & 16\\
HD\,148374 &     16 23 47.16 & +61 41 47.3 & $5.807 \pm 0.003$ & $152.16^{+0.86}_{-0.94}$   & $3.75^{+0.19}_{-0.04}$ & $6347^{+122}_{-24}$ & $2.67^{+0.03}_{-0.01}$ & 150 & 17\\
HD\,169487\,A &  18 22 11.80 & +51 32 30.9 & $6.818 \pm 0.003$ & $158.79^{+0.41}_{-0.51}$   & $2.75^{+0.04}_{-0.04}$ & $10291^{+62}_{-37}$ & $3.86^{+0.01}_{-0.01}$ & 300 & 11\\
57\,Cnc    &     08 54 14.73 & +30 34 44.8 & $5.774 \pm 0.003$ & $115.21^{+0.91}_{-0.85}$   & $2.80^{+0.04}_{-0.04}$ & $5269^{+272}_{-46}$ & $2.76^{+0.03}_{-0.04}$ & 150 & 10\\
$\gamma$\,And &  02 03 54.76 & +42 19 51.3 & $4.863 \pm 0.011$ & $85.40^{+6.09}_{-6.12}$    & $2.62^{+0.09}_{-0.09}$ & $9611^{+17}_{-16}$ & $4.02^{+0.03}_{-0.03}$ & 60 & 11\\
HD\,11031 &   01 49 15.53 & +47 53 48.9 & $6.458 \pm 0.003$ & $137.14^{+2.41}_{-2.12}$   & $2.72^{+0.05}_{-0.05}$ & $10407^{+59}_{-45}$ & $3.88^{+0.01}_{-0.01}$& 300 & 10\\
$\kappa$\,And &  23 40 24.51 & +44 20 02.2 & $4.148 \pm 0.003$ & $51.44^{+0.44}_{-0.60}$    & $2.80^{+0.04}_{-0.05}$ & $11101^{+35}_{-104}$ & $3.98^{+0.01}_{-0.01}$ & 60 & 14\\
$\lambda$\,Cas & 00 31 46.35 & +54 31 20.1 & $4.648 \pm 0.004$ & $137.76^{+13.61}_{-10.50}$ & $3.69^{+0.13}_{-0.12}$ & $11217^{+33}_{-36}$ & $3.46^{+0.02}_{-0.02}$ & 60 & 13\\
7\,CrB\,A &      15 39 22.67 & +36 38 08.9 & $4.985 \pm 0.003$ & $155.55^{+2.22}_{-2.66}$   & $3.89^{+0.05}_{-0.05}$ & $12157^{+34}_{-38}$ & $3.76^{+0.01}_{-0.01}$ & 60 & 30\\
7\,CrB\,B &      15 39 22.24 & +36 38 12.6 & $5.975 \pm 0.003$ & $155.33^{+0.88}_{-0.78}$   & $3.60^{+0.05}_{-0.05}$ & $13292^{+58}_{-74}$ & $4.04^{+0.01}_{-0.02}$ & 300 & 25\\
HD\,214007 &     22 33 53.15 & +61 46 41.3 & $6.535 \pm 0.003$ & $162.60^{+0.65}_{-0.47}$   & $2.37^{+0.04}_{-0.04}$ & $7907^{+9}_{-9}$ & $3.46^{+0.01}_{-0.01}$ & 300 & 16\\
$\iota$\,Her &   17 39 27.89 & +46 00 22.8 & $3.791 \pm 0.004$ & $154.11^{+6.07}_{-5.27}$   & $5.40^{+0.12}_{-0.08}$ & $14346^{+104}_{-96}$ & $3.61^{+0.02}_{-0.02}$ & 60 & 38\\
HD\,201433\,A &  21 08 38.89 & +30 12 20.3 & $5.719 \pm 0.003$ & $121.46^{+1.51}_{-1.34}$   & $3.09^{+0.05}_{-0.05}$ & $11786^{+63}_{-126}$ & $4.00^{+0.01}_{-0.01}$ & 150 & 17\\
\bottomrule
\end{tabular}}\label{tab:obs}
\end{table*}

\section{Method}

All targets listed in \mbox{Table\,\ref{tab:obs}\hspace{-1.5mm}} were observed with FLECHAS at the University Observatory Jena between 29 April and 6 September 2024. Three spectra were recorded per observation epoch, which were then combined into a fully reduced spectrum during data reduction. Before the actual spectroscopy of each target, three spectra were taken from a bulb lamp for flat-field correction and from a a thorium-argon lamp for wavelength-calibration, respectively. Three spectra of a bulb lamp were recorded before the spectroscopy of each target for flat-field correction, and three spectra of a thorium-argon lamp were taken for wavelength calibration. For dark subtraction, three darks were median combined into a master dark for each detector integration time. The signal-to-noise ratio (SNR) of all fully reduced spectra was measured at a wavelength of 6500\,\AA. The RV was then determined using the Doppler shift $\Delta\lambda=\lambda-\lambda_{0}$ of the H$\alpha$-line:
\begin{equation}
\text{RV} = c~\frac{\Delta\lambda}{\lambda_{0}}+BC
\end{equation}
where $\lambda$ is the measured wavelength of the spectral line, $\lambda_{0}$ is the rest wavelength, $c$ is the speed of light and $BC$ is the barycentric correction.

The wavelength of the H$\alpha$-line was determined by fitting a Gaussian profile on the Doppler-broadened core of the spectral line using the IRAF \citep{tody1993} task \texttt{splot}. The uncertainty of the RV measurements ($\Delta$RV), given below, results from the error in the Gaussian-fitting and in the wavelength calibration. This error can be assumed to be constant for all RV measurements of the same star since we always examined the H$\alpha$-line which does not spontaneously change in its width; thus, the error of the RV measurements only depends on the quality of the respective spectra, i.e. the signal-to-noise ratio (SNR). Only negligible variations in the SNR values were observed, justifying the use of a constant $\Delta$RV.

Those stars whose RV remained constant during the given period of observation are presented in Section\,\ref{sec:stable} of this paper. For illustration purposes, their RV is plotted over the barycentric Julian date (BJD), which is given in the format of \mbox{$\text{BJD} - 2460000$}.

Stars with variable RV are discussed in this paper in Section\,\ref{sec:variable}. In these cases, a Keplerian orbital solution was fitted to the RV data using the spectroscopic binary solver \citep{johnson2004}. The RV of a component of a binary system varies by:
\begin{equation}
\text{RV} = K~[e~\cos{(\omega)} + \cos{(\omega+\nu)}]+\gamma
\end{equation}
while the component orbits the barycenter of the system, where $K$ is the RV semi-amplitude, $\nu$ is the true anomaly of the component, $\omega$ is the argument of periastron, $e$ is the orbital eccentricity and $\gamma$ is the systemic velocity. $K$ is further given by:
\begin{equation}
K = \frac{2\pi~a\sin{(i)}}{P\sqrt{1-e^{2}}}
\end{equation}
with the minimum semi-major axis $a\sin{(i)}$ of the orbit of the component around the barycenter of the system and the orbital period $P$ of the system.

\section{Stars with Constant Radial Velocity}\label{sec:stable}

From the sample of stars shown in \mbox{Table\,\ref{tab:obs}\hspace{-1.5mm}}, eight were found to have a constant RV during our spectroscopic monitoring. These are: 17\,Dra\,A, HD\,148374, HD\,169487\,A, 57\,Cnc\,A, $\gamma$\,And, \mbox{HD\,11031,} $\kappa$\,And, and $\lambda$\,Cas. To determine how well the model of a constant RV matches the RV measurements of an observed target, considering the given measurement uncertainties, we calculated the reduced $\chi^2$-value for each fit, which is defined as follows:
\begin{equation} \label{eq:red_chi_squared}
\chi^{2}_{\text{red}} = \frac{1}{n-1}~\sum\left(\frac{RV - \overline{RV}}{\Delta RV}\right)^{2}
\end{equation}
Here, $n$ is the number of the individual measurements $RV$ with their typical uncertainty $\Delta RV $, and their mean value $\overline{RV}$.

In addition, a linear function was fitted to the RV measurements of each target to check whether a linear trend in the RV data over time can be detected. If the slope $\frac{\text{d}\text{RV}}{\text{dt}}$ of the fitted linear function is not significantly different from zero, this indicates that the RV of the target remained stable during the observation period. The significance of this slope was evaluated using Pearson's test. None of the eight stars presented in the following subsections were found to have a significant linear trend or periodic variation in their RV.

\subsection{17\,Dra\,A}

The Washington Visual Double Star Catalog \citep[WDS from here on, see][]{mason2001} lists three companions of 17\,Dra (see \mbox{Table\,\ref{tab:17Dra_wds}\hspace{-1.5mm}} for details). However, of these three stars, only the B and C components could be confirmed as part of the 17\,Dra system, as they all have very similar parallaxes and proper motions listed in the Gaia DR3. 17\,Dra\,A itself has a parallax of \mbox{$\varpi_{\text{A}}=7.89\pm0.10$\,mas} and a proper motion of \mbox{$\mu^{*}_{\alpha,\text{A}}=-12.24\pm0.13$\,mas/yr} and \mbox{$\mu_{\delta,\text{A}}=26.82\pm0.16$\,mas/yr}. The corresponding values of 17\,Dra\,B are and 17\,Dra\,C are: \mbox{$\varpi_{\text{B}}=7.52\pm0.12$\,mas}, \mbox{$\mu^{*}_{\alpha,\text{B}}=-15.54\pm0.17$\,mas/yr}, \mbox{$\mu_{\delta,\text{B}}=31.48\pm0.21$\,mas/yr}, and \mbox{$\varpi_{\text{C}}=7.74\pm0.09$\,mas}, \mbox{$\mu^{*}_{\alpha,\text{C}}=-12.78\pm0.12$\,mas/yr}, \mbox{$\mu_{\delta,\text{C}}=27.64\pm0.15$\,mas/yr}. In contrast, the D component from the WDS has a parallax of \mbox{$\varpi = 3.4\pm0.30$\,mas} and can therefore be excluded as part of the system. In addition to the confirmed WDS components, another companion  of 17\,Dra\,A was found in the Gaia DR3. This companion, henceforth referred to as 17\,Dra\,D, is located south of the star ($\theta\sim189\,^\circ$) at an angular separation of $\rho=80.7\,''$ (or $\sim10200$\,au projected separation). This companion has a parallax of \mbox{$\varpi=7.95\pm0.05$\,mas} and a proper motion of \mbox{$\mu^{*}_{\alpha}=-12.53\pm0.09$\,mas/yr} and \mbox{$\mu_{\delta}=27.13\pm0.10$\,mas/yr}, which agree well with the parallax and proper motion of 17\,Dar\,A. Thus, 17\,Dra is a hierarchical quadruple system consisting of two binary systems (with projected separations of 390 and 1500\,au) separated by about 11000\,au.

To quantify the degree of common proper motion (cpm) of all companions directly detected by Gaia, we calculated the cpm-index according as defined by \cite{mugrauer2019}. Additionally, we compared the detected differential proper motion between each companion relative to its primary star with the companion's escape velocity, as estimated by \cite{mugrauer2019}. In the case of the 17\,Dra system, the companions detected have cpm-indices of 11, 61 and 138 for 17\,Dra\,B, C, and D respectively, meaning they all exhibit a high degree of cpm with the primary star of the system. Furthermore, the parallaxes of the components do not significantly deviate from each other, and their differential proper motions relative to 17\,Dra\,A do not exceed the estimated escape velocities. This is all as expected for components of a gravitationally bound multiple star system.

\begin{center}
\begin{table}[h!]
\caption{Components of 17\,Dra, listed in the WDS, with their angular separation $\rho$ and position angle $\theta$, measured in the first and last observing epoch.}
\centering
\resizebox{\hsize}{!}{
\begin{tabular}{ccccccc}
\toprule
Comp. & Date 1 & $\rho_1$ [$''$]  & $\theta_1$ [$\deg$] &  Date 2  & $\rho_2$ [$''$] & $\theta_2$ [$\deg$]  \\
\midrule
B  & 1781 & 4.0 &  114 &  2022 & 3.0 & 104  \\
C  & 1823 & 90.3 & 196 & 2018 & 90.2 & 193  \\
CD & 1879 & 116.9 & 122 & 2020 & 125.5 & 123  \\
\bottomrule
\end{tabular}}\label{tab:17Dra_wds}
\end{table}
\end{center}

\begin{table}[h!]
\caption{For 17\,Dra\,A, 16 RV measurements were carried out over a period of \mbox{79 days}. The typical uncertainty of the RV measurements is \mbox{$\Delta\text{RV} = 1.20$\,km/s}. The table includes the SNR of all spectra.}
\centering
\begin{tabular}{ccc}
\toprule
$\text{BJD}-2460000$ & RV [km/s] & SNR\\ \midrule
430.48932 & $-14.47$ &  180\\
432.41889 & $-12.95$ &  164\\
439.45953 & $-14.92$ &  187\\
444.44487 & $-13.55$ &  176\\
445.43250 & $-15.14$ &  162\\
446.38934 & $-14.51$ &  128\\
451.37339 & $-15.86$ &  172\\
468.48837 & $-16.87$ &  170\\
469.50364 & $-14.86$ &  161\\
473.52038 & $-14.49$ &  137\\
481.49545 & $-15.93$ &  139\\
486.39842 & $-13.49$ &  191\\
487.39599 & $-13.13$ &  177\\
501.47496 & $-16.70$ &  179\\
508.47437 & $-14.25$ &  145\\
509.51423 & $-15.87$ &  122\\
\bottomrule
\end{tabular}\label{tab:17Dra}
\end{table}

\begin{figure}[h!]
\centering
\includegraphics[width=1\linewidth]{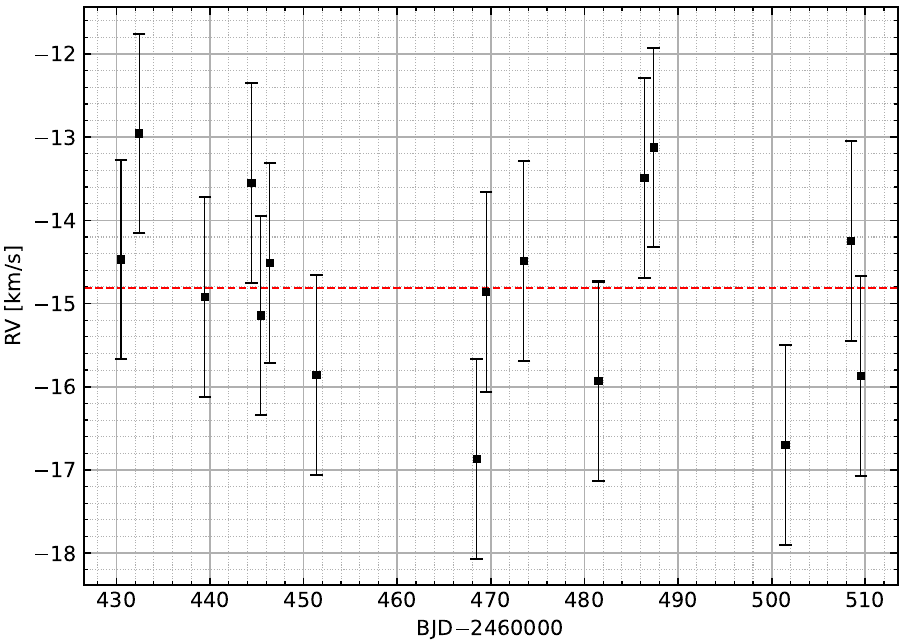}
\caption{The RV of 17\,Dra\,A measured over a period of \mbox{79 days}. The mean RV is indicated by the dashed red line.}
\label{fig:17Dra}
\end{figure}

Spectroscopic observations of 17\,Dra\,A were performed with FLECHAS in 16 observation epochs over a period of \mbox{79 days}. The obtained individual RV measurements of the star are summarized in \mbox{Table\,\ref{tab:17Dra}\hspace{-1.5mm}}. As illustrated in \mbox{Figure\,\ref{fig:17Dra}\hspace{-1.5mm}}, the RV of the star is constant during the entire spectroscopic observation time.

\subsection{HD\,148374}

The WDS lists \textbf{two} components for HD\,148374, which are summarized in \mbox{Table\,\ref{tab:HD148374_wds}\hspace{-1.5mm}}. The Ab component is not detected by Gaia due to its small angular separation of only 0.2\,$''$ (or about 30\,au projected separation), as expected from the direct imaging performance of Gaia, determined by \cite{mugrauer2025}. In contrast, an object is listed near the position of the B component in the Gaia DR3 ($\rho\sim0.858$\,$''$ and $\theta\sim359.1\,^\circ$), but neither parallax nor proper motion are given for this source. If this object is the same source that is listed for the first observation epoch in the WDS, due to the Gaia DR3 proper motion of HD\,148374 (\mbox{$\mu^{*}_{\alpha,\text{A}}=-29. 26\pm0.051$\,mas/yr} and \mbox{$\mu_{\delta,\text{A}}=35.25\pm0.06$\,mas/yr}) and the long epoch difference of \mbox{184 years} between the first WDS and the Gaia observation (reference epoch 2016.0) this object must be a true companion of the star. If this source discovered in 1832 were a background star, it would have to be located at $\rho\sim7.8$\,$''$ and $\theta\sim136\,^\circ$ in 2016, i.e. at a significantly different position than listed in the Gaia DR3. In contrast, the observed change in position angle of 8\,$^\circ$ between the first WDS and the Gaia observation epoch is well compatible with the orbital motion in \mbox{184 years} expected for such a companion at a projected separation of about 140\,au. Since the status of the nearby component Ab is not yet clear, we classify HD\,148374 as the main component of a binary star system.

\begin{table}[h!]
\caption{Components of HD\,148374, listed in the WDS, with their angular separation $\rho$ and position angle $\theta$, measured in the first and last observing epoch.}
\centering
\resizebox{\hsize}{!}{\begin{tabular}{ccccccc}
\toprule
Comp. &  Date 1 &$\rho_1$ [$''$]  & $\theta_1$ [$\deg$]  & Date 2 & $\rho_2$ [$''$] & $\theta_2$ [$\deg$]   \\
\midrule
Ab & 1986 & 0.2 & 174 &  1990 & 0.2 & 154   \\
B & 1832 & 0.9 & 7 &  2021 & 0.9 & 350  \\
\bottomrule
\end{tabular}} \label{tab:HD148374_wds}
\end{table}

\begin{figure}[h!]
\centering
\includegraphics[width=\linewidth]{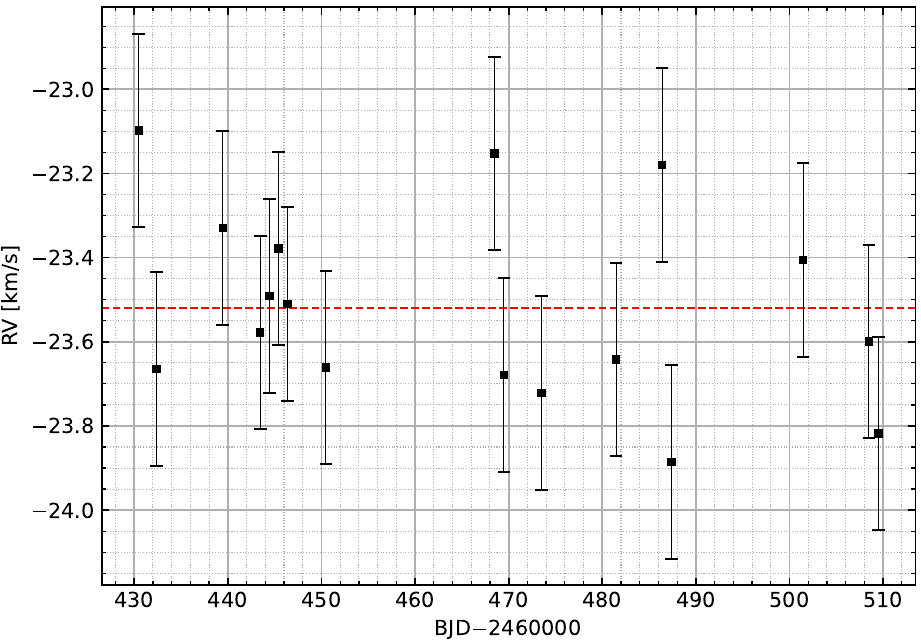}
\caption{The RV of HD\,148374 measured over a period of \mbox{79 days}. The mean RV is indicated by the dashed red line.}\label{fig:HD148374}
\end{figure}

A total of 17 RV measurements of HD\,148374 were carried out with FLECHAS over a period of \mbox{79 days}. All these measurements are listed in \mbox{Table\,\ref{tab:HD148374}\hspace{-1.5mm}} and plotted in \mbox{Figure\, \ref{fig:HD148374}\hspace{-1.5mm}}. No significant RV variation of the star was detected during our entire spectroscopic monitoring program of the star.

\begin{table}[h!]
\caption{For HD\,148374, 17 RV measurements were carried out over a period of \mbox{79 days}. The typical error of each RV measurement is \mbox{$\Delta\text{RV} = 0.23$\,km/s}. The table includes the SNR of all spectra.}
\centering
\begin{tabular}{ccc}
\toprule
$\text{BJD}-2460000$ & RV [km/s] &  SNR\\ \midrule
430.47855   & $-23.10$ &  192 \\
432.38072   & $-23.66$ &  177 \\
439.45192   & $-23.33$ &  197 \\
443.46678   & $-23.58$ &  139 \\
444.43428   & $-23.49$ &  172 \\
445.42352   & $-23.38$ &  180 \\
446.38157   & $-23.51$ &  131 \\
450.47754   & $-23.66$ &  172 \\
468.48009   & $-23.15$ &  191 \\
469.49589   & $-23.68$ &  181 \\
473.51291   & $-23.72$ &  160 \\
481.48415   & $-23.64$ &  224 \\
486.38958   & $-23.18$ &  206 \\
487.38728   & $-23.89$ &  197 \\
501.46656   & $-23.41$ &  192 \\
508.46658   & $-23.60$ &  152\\
509.50457   & $-23.82$ &  158 \\
\bottomrule
\end{tabular}\label{tab:HD148374}
\end{table}

\subsection{HD\,169487\,A}

HD\,169487 has no entries in the WDS. However, the star has a very wide ($\rho\sim315\,''$ or about 50200\,au projected separation) companion in the Gaia DR3, which is about 9\,mag fainter in the G-band than HD\,169487. Therefore, we classify HD\,169487 as the primary component of a wide binary system and will therefore be referred to as HD\,169487\,A from here on. HD\,169487\,A has a parallax of \mbox{$\varpi_{\text{A}}=6.2801\pm0.0203$\,mas} and a proper motion of \mbox{$\mu^{*}_{\alpha,\text{A}}=15.905\pm0.025$\,mas/yr} and \mbox{$\mu_{\delta,\text{A}}=-12.879\pm0.027$\,mas/yr}. The detected companion has a parallax of \mbox{$\varpi_{\text{B}}=6.2809\pm0.0253$\,mas} and a proper motion of \mbox{$\mu^{*}_{\alpha,\text{B}}=16.337\pm0.032$\,mas/yr} and \mbox{$\mu_{\delta,\text{B}}=-12.247\pm0.029$\,mas/yr}. Gaia DR3 astrometry proves that HD\,169487\,A and B are equidistant, as their parallaxes do not significantly deviate from each other. The companion has a cpm-index of 53, meaning the two components of this star system exhibit a high degree of common proper motion. However, it should be noted that the companion's differential proper motion to its primary is approximately 1.6 times greater than its estimated escape velocity. Since it is very unlikely that a binary star system would be observed during a destructive event, the higher differential proper motion could suggest the presence of another close companion around one of the components of the system. Such a close companion with its rapid orbital motion leads to a shift in the star's photocenter, which can induce an increased differential proper motion.

\begin{figure}[h!]
\centering
\includegraphics[width=1\linewidth]{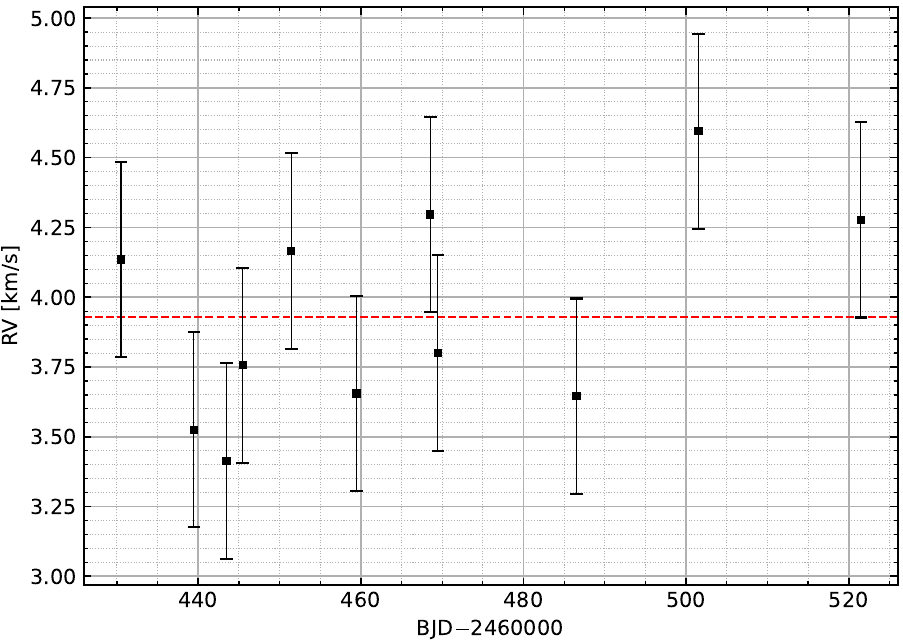}
\caption{The RV of HD\,169487\,A measured over a period of \mbox{91 days}. The mean RV is indicated by the dashed red line.}\label{fig:HD169487}
\end{figure}

The RV of HD\,169487\,A was monitored with FLECHAS in 11 observation epochs over a period of \mbox{91 days}. The RV of the star was constant during our entire spectroscopic observation campaign, as shown in \mbox{Table\,\ref{tab:HD169487}\hspace{-1.5mm}} and \mbox{Figure\,\ref{fig:HD169487}\hspace{-1.5mm}}.

\begin{table}[h!]
\caption{For HD\,169487\,A, 11 RV measurements were carried out over a period of \mbox{91 days}. The typical error of each RV measurement is $\Delta\text{RV} = 0.35$\,km/s. The table includes the SNR of all spectra.}
\centering
\begin{tabular}{ccc}
\toprule
$\text{BJD}-2460000$ & RV [km/s] &  SNR\\ \midrule
430.50341  & $4.14$ &  117\\
439.47117  & $3.53$ &  126\\
443.48021  & $3.41$ &  95\\
445.46825  & $3.76$ &  112\\
451.44839  & $4.17$ &  132\\
459.44927  & $3.66$ &  95\\
468.53879  & $4.30$ &  120\\
469.47046  & $3.80$ &  116\\
486.50704  & $3.65$ &  139\\
501.52976  & $4.59$ &  131\\
521.47611  & $4.28$ &  99\\
\bottomrule
\end{tabular} \label{tab:HD169487}
\end{table}

\subsection{57\,Cnc}

57\,Cnc is listed in the WDS as a hierarchical triple star system (see \mbox{Table\,\ref{tab:57Cnc_wds}\hspace{-1.5mm}}).
\begin{table}[h!]
\caption{Components of 57\,Cnc, listed in the WDS, with their angular separation $\rho$ and position angle $\theta$, measured in the first and last observing epoch.}
\centering
\resizebox{\hsize}{!}{\begin{tabular}{ccccccc}
\toprule
Comp. &  Date 1 & $\rho_1$ [$''$]  & $\theta_1$ [$\deg$]  &  Date 2 & $\rho_2$ [$''$] & $\theta_2$ [$\deg$] \\ \midrule
B & 1782 & 0.9 & 338 & 2019 & 1.5 & 310  \\
AB, C & 1921 & 55.8 & 198 & 2015 & 54.6 & 204  \\
\bottomrule
\end{tabular}}\label{tab:57Cnc_wds}
\end{table}

However, only the B component can be confirmed as a true companion of the star, as it has a Gaia DR3 parallax and proper motion similar to those of 57\,Cnc\,A:\newline \mbox{$\varpi_{\text{A}}=8.64\pm0.07$\,mas}, \mbox{$\varpi_{\text{B}}=8.61\pm0.06$\,mas}\newline
\mbox{$\mu^{*}_{\alpha,\text{A}}=41.21\pm0.08$\,mas/yr}, \mbox{$\mu_{\delta,\text{A}}=-23.25\pm0. 06$\,mas/yr}\newline \mbox{$\mu^{*}_{\alpha,\text{B}}=37.55\pm0.05$\,mas/yr}, \mbox{$\mu_{\delta,\text{B}}=-24.89\pm0.06$\,mas/yr}. The companion has a cpm-index of 23, hence is clearly co-moving with 57\,Cnc\,A. In addition, the differential proper motion of the companion to 57\,Cnc\,A is smaller than its estimated escape velocity as expected for a gravitationally bound companion. \newline In contrast, the C component listed in the WDS has a cpm index of only 0.5, and hence can clearly be ruled out as a co-moving companion of 57\,Cnc\,A. Furthermore, the Gaia DR3 parallax $\varpi=1.96\pm0.02$\,mas of this star is very different from that of 57\,Cnc\,A and B, which clearly disqualifies it as part of the 57\,Cnc system. Since no other companions of the star were found in the Gaia DR3, 57\,Cnc is a binary system with a projected separation of about 180\,au and not a triple star system as suggested in the WDS.

\begin{figure}[h!]
\centering
\includegraphics[width=1\linewidth]{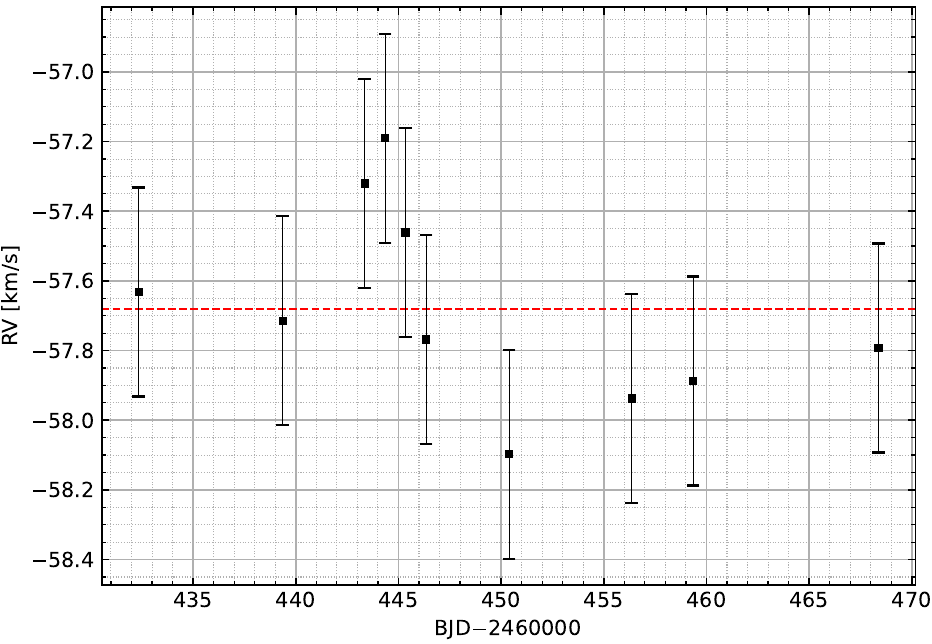}
\caption{The RV of 57\,Cnc measured over a period of \mbox{36 days}. The mean RV is indicated by the dashed red line.}\label{fig:57Cnc}
\end{figure}

The RV of 57\,Cnc, was measured in our spectroscopic monitoring program in ten observation epochs over a period of \mbox{36 days} and proved to be stable within this span of time. All individual RV measurements of the star are listed in \mbox{Table\,\ref{tab:57Cnc}\hspace{-1.5mm}} and plotted in \mbox{Figure\,\ref{fig:57Cnc}\hspace{-1.5mm}}.

\begin{table}[h!]
\caption{For 57\,Cnc ten RV measurements were carried out over a period of \mbox{36 days}. The typical error of each RV measurement is $\Delta\text{RV} = 0.30$\,km/s. The table includes the SNR of all spectra.}
\centering
\begin{tabular}{ccc}
\toprule
$\text{BJD}-2460000$ & RV [km/s] &  SNR\\ \midrule
432.35093  & $-57.63$ &  177\\
439.36150  & $-57.71$ &  187\\
443.35848  & $-57.32$ &  172\\
444.34824  & $-57.19$ &  179\\
445.34874  & $-57.46$ &  180\\
446.34396  & $-57.77$ &  128\\
450.39132  & $-58.10$ &  150\\
456.36031  & $-57.94$ &  156\\
459.35548  & $-57.89$ &  147\\
468.37627  & $-57.79$ &  126\\
\bottomrule
\end{tabular}\label{tab:57Cnc}
\end{table}

\subsection{$\gamma$\,And}

For $\gamma$\,And three components are listed in the WDS (see \mbox{Table\,\ref{tab:gamAnd_wds}\hspace{-1.5mm}}) but none of these stars are detected by Gaia. No other companions of the star are listed in the Gaia DR3. Since the angular separation and position angle of the BC component has not changed much with \mbox{244 years} of epoch difference between the first and the last WDS observation, we conclude that it is a true companion of $\gamma$\,And with a projected separation to the star of about 1000\,au.

If this close pair of stars discovered in 1777 were a background source with negligible proper motion in 2021 it would have to be located at $\rho\sim18\,''$ and $\theta\sim23\,^\circ$ according to the Gaia DR3 proper motion of $\gamma$\,And (\mbox{$\mu^{*}_{\alpha,\text{A}}=16.645\pm0.930$\,mas/yr}, \mbox{$\mu_{\delta,\text{A}}=-47.746\pm0.90$\,mas/yr}) and the long epoch difference of \mbox{244 years}, but this differs significantly from the position, measured in the last WDS observation epoch. For this close binary system an orbital period of about \mbox{63 years} was derived, as reported in the 6th Catalog of Orbits of Visual Double Stars \citep{hartkopf2001}. Furthermore, according to the WDS, the B component is itself a spectroscopic binary with a period of about \mbox{2.7 days}. In contrast, the status of the D component listed in the WDS is ambiguous, as its detection was reported for only one observation epoch, and no source is listed near its position in the Gaia DR3. From this we conclude that $\gamma$\,And is the primary component of a hierarchical quadruple system.

\begin{table}[h!]
\caption{Components of $\gamma$\,And, listed in the WDS, with their angular separation $\rho$ and position angle $\theta$, measured in the first and last observing epoch.}
\centering
\resizebox{\hsize}{!}{\begin{tabular}{ccccccc}
\toprule
Comp. &  Date 1 & $\rho_1$ [$''$]  & $\theta_1$ [$\deg$]  &  Date 2 & $\rho_2$ [$''$] & $\theta_2$ [$\deg$]   \\ \midrule
BC & 1777 & 12.0 & 67 & 2021 & 9.5 & 63  \\
D & 1898 & 27.9 & 245 & --   & -- & -- \\
\bottomrule
\end{tabular}}\label{tab:gamAnd_wds}
\end{table}

$\gamma$\,And was observed with FLECHAS in 11 observation epochs within a monitoring period of \mbox{44 days}, but no significant variation in its RV was detected.

\begin{table}[h!]
\caption{For $\gamma$\,And, eleven RV measurements were carried out over a period of \mbox{44 days}. The typical error of each RV measurement is $\Delta\text{RV} = 0.60$\,km/s. The table includes the SNR of all spectra.}
\centering
\begin{tabular}{ccc}
\toprule
$\text{BJD}-2460000$ & RV [km/s] &  SNR\\ \midrule
516.56282  & $-8.55$ &  536\\
520.54226  & $-8.32$ &  715\\
531.52621  & $-7.90$ &  606\\
533.46280  & $-7.88$ &  680\\
535.57245  & $-7.10$ &  782\\
538.51730  & $-7.55$ &  700\\
545.50779  & $-7.30$ &  552\\
549.56800  & $-6.78$ &  645\\
550.54184  & $-6.58$ &  729\\
555.44836  & $-7.53$ &  625\\
560.50345  & $-8.11$ &  521\\
\bottomrule
\end{tabular}\label{tab:gamAnd}
\end{table}

\begin{figure}[h!]
\centering
\includegraphics[width=1\linewidth]{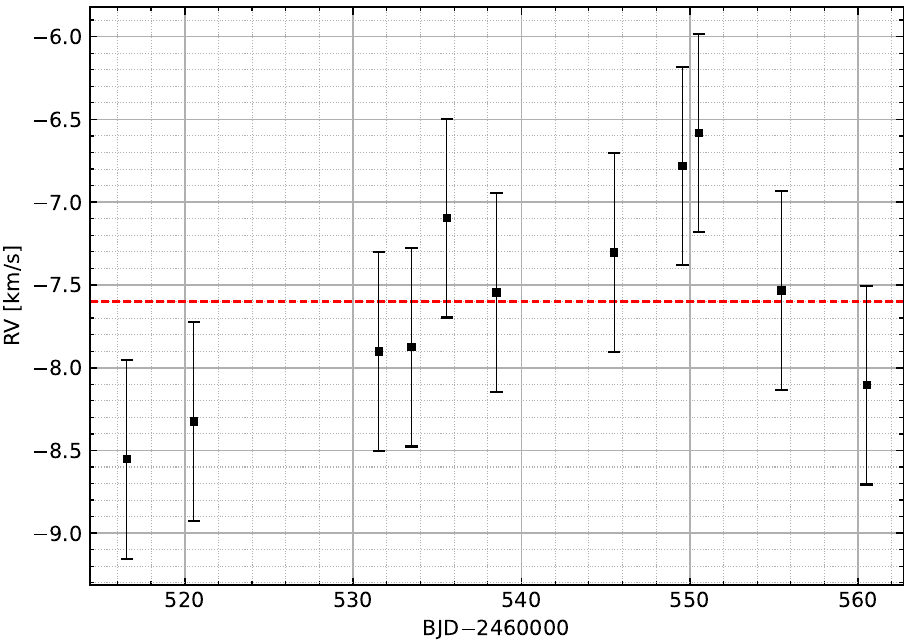}
\caption{The RV of $\gamma$\,And measured over a period of \mbox{44 days}. The mean RV is indicated by the dashed red line.}\label{fig:gamAnd}
\end{figure}

\subsection{HD\,11031}

HD\,11031 is listed in the WDS as a multiple star system (see \mbox{Table\,\ref{tab:HD11031_wds}\hspace{-1.5mm}}). The companionship of the B and C components to HD\,11031\,A can be confirmed with their Gaia DR3 astrometry due to comparable parallaxes
(\mbox{$\varpi_{\text{A}}=7.29\pm0.12$\,mas}, \mbox{$\varpi_{\text{B}}=6.44\pm0.17$\,mas}, \mbox{$\varpi_{\text{C}}=7.76\pm0.03$\,mas}) and proper motions:\newline \mbox{$\mu^{*}_{\alpha,\text{A}}=-16.92\pm0.15$\,mas/yr}, \mbox{$\mu_{\delta,\text{A}}=-2.85\pm0.10$\,mas/yr}\newline \mbox{$\mu^{*}_{\alpha,\text{B}}=-15.45\pm0.17$\,mas/yr}, \mbox{$\mu_{\delta,\text{B}}=-9.18\pm0.10$\,mas/yr} \newline \mbox{$\mu^{*}_{\alpha,\text{C}}=-20.38\pm0.04$\,mas/yr}, \mbox{$\mu_{\delta,\text{C}}=-9.53\pm0.03$\,mas/yr}.\newline In contrast, the D component listed in the WDS can clearly be excluded as a companion of the star, as it has a cpm-index of only 0.5 and its Gaia DR3 parallax (\mbox{$\varpi=2.89\pm0.03$\,mas}) is significantly different from that of HD\,11031\,A. Based on the Gaia DR3 astrometry, the projected separation between HD\,11031\,A \& B is about 270\,au, and $\sim$2800\,au between HD\,11031\,A \& C. Due to its small angular separation of only 0.1\,$''$, the Ab component is not listed in the Gaia DR3, but its companionship to HD\,11031 is proven, as an orbital solution with a period of about 37\,yr could be derived for this star, as reported in the WDS. This catalog also mentions that HD\,11031\,B itself is a spectroscopic double star, i.e. HD\,11031 is actually a quintuple star system. Components B and C have cpm-indices of about 5, meaning they share a common proper motion with HD\,11031\,A. However, the 5-parameter astrometric solution of the individual components of the HD\,11031 system show large astrometric excess noises of 0.836, 0.897, and 0.198\,mas for HD\,11031\,A, B, and C. Taking these noises into account, the parallaxes of the components agree with each other; that is to say, these stars are equidistant, as would be expected of gravitationally bound members of a star system. While the differential proper motion of HD\,11031\,B to the primary star is smaller than its estimated escape velocity, that of HD\,11031\,C is about 3.1 greater than the escape velocity estimated for this star. As described above, this high differential proper motion could be caused by a close companion of the star that has not been detected. In this context, it is worth noting that all components of the HD\,11031 system also have a high Renormalized Unit Weight Error (RUWE): 4.12 for HD\,11031\,A, 4.56  for HD\,11031\,B, and 1.78 for HD\,11031\,C. This suggests the possible presence of additional close companions of these stars. While such companions are already known orbiting HD\,11031\,A and HD\,11031\,B, $\text{RUWE}=1.78$ for HD\,11031\,C could indicate that also this star might harbor an additional close companion \citep[see e.g.][and references therein]{castro2024}, which would also explain the higher differential proper motion observed for this star.

\begin{table}[h!]
\caption{Components of HD\,11031, listed in the WDS, with their angular separation $\rho$ and position angle $\theta$, measured in the first and last observing epoch.}
\centering
\resizebox{\hsize}{!}{\begin{tabular}{ccccccc}
\toprule
Comp. & Date 1 & $\rho_1$ [$''$]  & $\theta_1$ [$\deg$] & Date 2&  $\rho_{2}$ [$''$] & $\theta_{2}$ [$\deg$]  \\ \midrule
Ab & 1984 & 0.1 & 194 & 1994 & 0.1 & 251  \\
B & 1828 & 2.2 & 227 & 2020 & 2.0 & 197  \\
C & 1828 & 20.6 & 177 & 2020 & 20.7 & 178  \\
D & 1881 & 134.7 & 95 & 2016 & 138.8 & 96  \\
\bottomrule
\end{tabular}}\label{tab:HD11031_wds}
\end{table}

\begin{figure}[h!]
\centering
\includegraphics[width=1\linewidth]{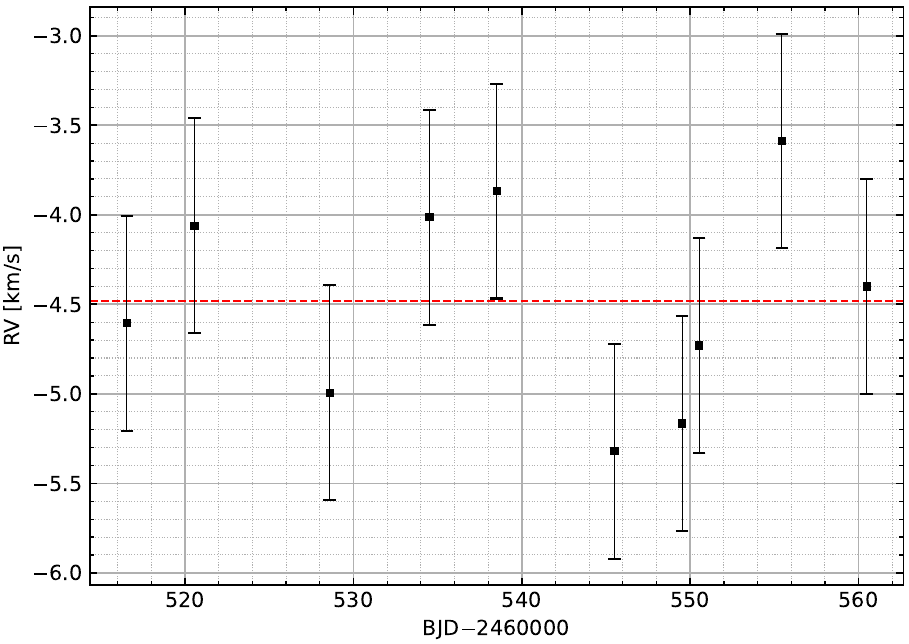}
\caption{The RV of HD\,11031 measured over a period of \mbox{44 days}. The mean RV is indicated by the dashed red line.}\label{fig:HD11031}
\end{figure}

The RV of HD\,11031 was monitored over a period of \mbox{44 days} in ten observing epochs, but no significant RV variation was detected. The RV measurements of the star are summarized in \mbox{Table\,\ref{tab:HD11031}\hspace{-1.5mm}} and are plotted in \mbox{Figure\,\ref{fig:HD11031}\hspace{-1.5mm}}.

\begin{table}[h!]
\caption{For HD\,11031, ten RV measurements were carried out over a period of \mbox{44 days}. The typical error of each RV measurement is \mbox{$\Delta\text{RV}=0.60$\,km/s}. The table includes the SNR of all spectra.}
\centering
\begin{tabular}{ccc}
\toprule
$\text{BJD}-2460000$ & RV [km/s] &  SNR \\ \midrule
516.54405  & $-4.61$ & 141 \\
520.56169  & $-4.06$ &  162 \\
528.58723  & $-4.99$ &  151 \\
534.51877  & $-4.01$ &  151 \\
538.50813  & $-3.87$ &  179 \\
545.49812  & $-5.32$ &  138 \\
549.51387  & $-5.16$ &  164 \\
550.53126  & $-4.73$ & 187 \\
555.43879  & $-3.59$ &  152 \\
560.48008  & $-4.40$ &  142 \\
\bottomrule
\end{tabular}\label{tab:HD11031}
\end{table}

\subsection{$\kappa$ And}

The WDS lists multiple components for $\kappa$\,And, which are summarized in \mbox{Table\,\ref{tab:kapAnd_wds}\hspace{-1.5mm}}. However, components B and C can be ruled out as real companions of the star, since both stars only have small cpm-indices of about 0.9 and their Gaia DR3 parallaxes (\mbox{$\varpi< 1.0~\text{mas}$}) differ greatly from that of $\kappa$\,And (\mbox{$\varpi=19.41\pm0.21$\,mas}). In contrast, the companion listed at an angular separation of about 1\,$''$ (or about 50\,au of projected separation) is of substellar nature and is in fact part of the system \citep{carson2013}. No additional companions of the star are detected in the Gaia DR3.

\begin{table}[h!]
\caption{Components of $\kappa$\,And, listed in the WDS, with their angular separation $\rho$ and position angle $\theta$, measured in the first and last observing epoch.}
\centering
\resizebox{\hsize}{!}{\begin{tabular}{ccccccc}
\toprule
Comp. & Date 1 & $\rho_1$ [$''$]  & $\theta_1$ [$\deg$] & Date 2 &  $\rho_2$ [$''$] & $\theta_2$ [$\deg$]  \\ \midrule
Ab & 2011 & 1.1 & 56 & 2012 & 1.0 & 51  \\
B & 1828 & 35.0 & 182 & 2002 & 47.4 & 202  \\
C & 1836 & 98.5 & 296 & 2012 & 115.3 & 293 \\
\bottomrule
\end{tabular}}\label{tab:kapAnd_wds}
\end{table}

\begin{figure}[h!]
\centering
\includegraphics[width=0.95\linewidth]{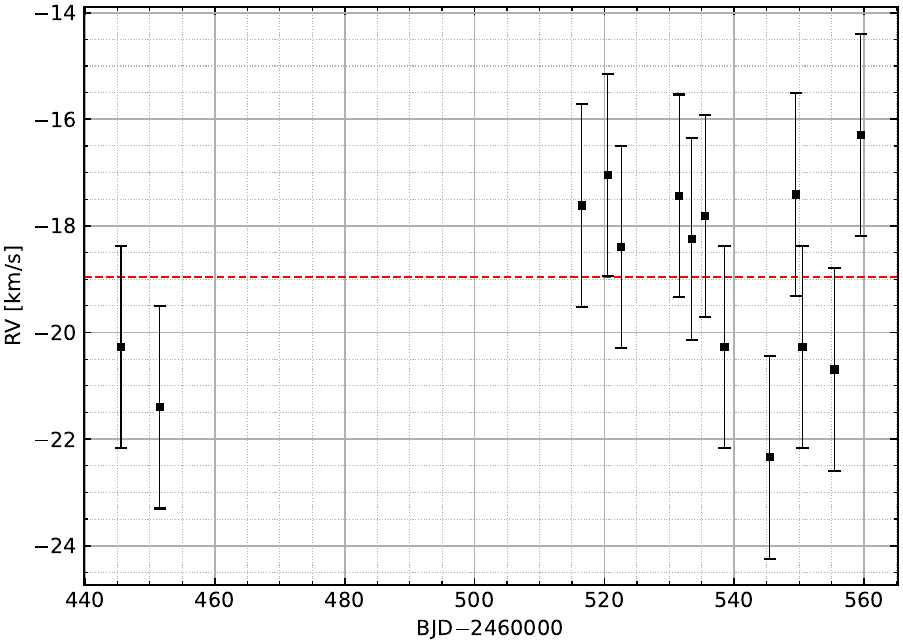}
\caption{The RV of $\kappa$\,And measured over a period of \mbox{114 days}. The mean RV is indicated by the dashed red line.}\label{fig:kapAnd}
\end{figure}

$\kappa$\,And was observed with FLECHAS in 14 observing epochs distributed over a period of \mbox{114 days}, during which time the RV of the star remained stable. All measurements are displayed in \mbox{Table\,\ref{tab:kapAnd}\hspace{-1.5mm}} and plotted in \mbox{Figure\,\ref{fig:kapAnd}\hspace{-1.5mm}}.

\begin{table}[h!]
\caption{For $\kappa$\,And, 14 RV measurements were carried out over a period of \mbox{114 days}. The typical error of each RV measurement is \mbox{$\Delta\text{RV} = 1.90$\,km/s}. The table includes the SNR of all spectra.}
\centering
\begin{tabular}{ccc}
\toprule
$\text{BJD}-2460000$ & RV [km/s] & SNR \\ \midrule
445.55471  & $-20.27$ &  124 \\
451.52514  & $-21.40$ &  163\\
516.53550  & $-17.62$ &  253\\
520.53391  & $-17.04$ &  332\\
522.56758  & $-18.39$ &  211\\
531.51545  & $-17.43$ &  256\\
533.45423  & $-18.25$ &  215\\
535.50917  & $-17.81$ &  215\\
538.49901  & $-20.27$ &  198\\
545.48871  & $-22.34$ &  174\\
549.45990  & $-17.41$ &  177\\
550.47650  & $-20.27$ &  72\\
555.42770  & $-20.69$ &  303\\
559.47165  & $-16.29$ &  113\\
\bottomrule
\end{tabular}\label{tab:kapAnd}
\end{table}

\subsection{$\lambda$\,Cas}

According to the WDS, $\lambda$\,Cas is a binary system (see \mbox{Table\,\ref{tab:lamCas_wds}\hspace{-1.5mm}}) for which a premature orbital solution with a period of about \mbox{246 years} was determined by \cite{drummond2014}.

\begin{table}[!h]
\caption{Components of $\lambda$\,Cas, listed in the WDS, with their angular separation $\rho$ and position angle $\theta$, measured in the first and last observing epoch.}
\centering
\resizebox{\hsize}{!}{\begin{tabular}{ccccccc}
\toprule
Comp. & Date 1 & $\rho_1$ [$''$]  & $\theta_1$ [$\deg$]  & Date 2 & $\rho_2$ [$''$] & $\theta_2$ [$\deg$]  \\ \midrule
B & 1845 & 0.3 & 122 & 2020 & 0.2 & 239  \\
\bottomrule
\end{tabular}}\label{tab:lamCas_wds}
\end{table}

\begin{figure}[h!]
\centering
\includegraphics[width=1\linewidth]{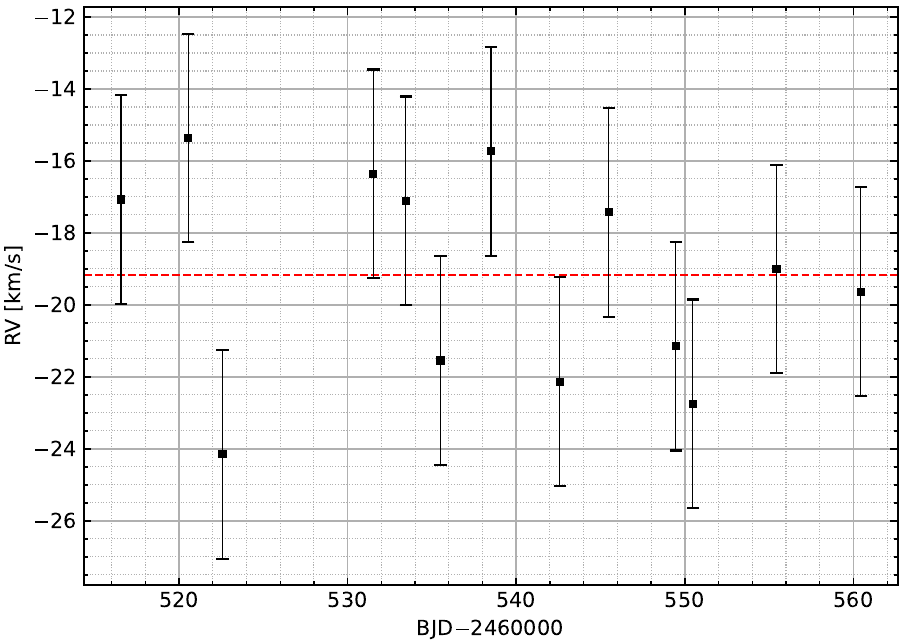}
\caption{The RV of $\lambda$\,Cas measured over a period of \mbox{44 days}. The mean RV is indicated by the dashed red line.}\label{fig:lamCas}
\end{figure}

A total of 13 RV measurements were carried out for $\lambda$\,Cas with FLECHAS over a period of \mbox{44 days}. The star's individual RV measurements are listed in \mbox{Table\,\ref{tab:lamCas}\hspace{-1.5mm}} and shown in \mbox{Figure\,\ref{fig:lamCas}\hspace{-1.5mm}}. No significant RV variability was observed during the entire observation period.

\begin{table}[h!]
\caption{For $\lambda$\,Cas, 13 RV measurements were carried out over a period of \mbox{44 days}. The typical error of each RV measurement is $\Delta\text{RV} = 2.90$\,km/s. The table includes the SNR of all spectra.}
\centering
\begin{tabular}{ccc}
\toprule
$\text{BJD}-2460000$ & RV [km/s] &  SNR \\\midrule
516.55504  & $-17.07$  &  111\\
520.53863  & $-15.36$  &  156\\
522.57169  & $-24.15$  &  160\\
531.52078  & $-16.36$  &  224\\
533.45677  & $-17.11$  &  151\\
535.51261  & $-21.54$  &  163\\
538.50164  & $-15.73$  &  148\\
542.58402  & $-22.13$  &  154\\
545.49174  & $-17.43$  &  132\\
549.46305  & $-21.14$  &  123\\
550.47951  & $-22.75$  &  141\\
555.43250  & $-19.00$  &  145\\
560.43533  & $-19.63$  &  129\\
\bottomrule
\end{tabular}\label{tab:lamCas}
\end{table}

\subsection{Summary}\label{sec:summary_const}

\mbox{Table\,\ref{tab:stable}\hspace{-1.5mm}} summarizes the results for the targets presented in the previous subsections. All of these stars showed a constant RV during our spectroscopic monitoring program, which was performed with FLECHAS at the University Observatory Jena. The table lists the mean and standard deviation of the RV measurements for each star, as well as the $\chi^2_\text{red}$-value of the fits to the RV data for these stars, assuming $\text{RV}=\text{const}$. As the $\chi^2_\text{red}$-value of all fits is around one, the RV measurements of all targets are consistent with a constant RV within their uncertainties.

\begin{table}[h!]
\caption{List of all stars that have a constant RV in our spectroscopic monitoring program with FLECHAS. For each target, we list the mean and standard deviation of its RV, as well as the $\chi^2_\text{red}$-value of the fit assuming a constant RV.}
\centering
\begin{tabular}{ccc}
\toprule
Target &  $\overline{\text{RV}}$ [km/s] &  $\chi^{2}_\text{red}$ \\ \midrule
17\,Dra\,A & $-14.81\pm1.20$ &  1.00\\
HD\,148374 & $-23.52\pm0.23$ &  1.01 \\
HD\,169487\,A & $3.93\pm0.39$ &  1.17 \\
57\,Cnc & $-57.68\pm0.28$ &  0.90 \\
$\gamma$\,And & $-7.60\pm0.63$ & 1.09 \\
HD\,11031 & $-4.48\pm0.59$ &  0.95 \\
$\kappa$\,And & $-18.96\pm1.86$ &  0.96 \\
$\lambda$\,Cas & $-19.18\pm2.92$ &  1.01 \\
\bottomrule
\end{tabular}\label{tab:stable}
\end{table}

\section{Stars with Variable Radial Velocity}\label{sec:variable}

Of the total sample of targets shown in \mbox{Table\,\ref{tab:obs}\hspace{-1.5mm}}, five stars exhibit significant variability in their RV measurements. Keplerian orbital solutions were fitted to the RV data for all these targets to determine the orbital elements of these spectroscopic binary systems. In order to accurately determine these elements, it is important to obtain a sufficiently large number of RV measurements that cover the system's orbital phase adequately. The $\chi^2_{\text{red}}$-value is given for all best-fitting Keplerian orbital solutions, as a measure of how well these solutions agree with the RV measurements, taking into account their uncertainties. All targets, with variable RV, are discussed in more detail in separate subsections below.

For single-lined spectroscopic binaries (SB1) their mass function:
\begin{equation}\label{eq:mass_orbit}
    f(m) = \frac{\left(mass_2~\sin{i}\right)^3}{\left(mass_1 + mass_2\right)^{2}}
    = \frac{K^{3}~P~(1-e^{2})^{3/2}}{2\pi G}
\end{equation}
was determined, where $mass_1$ is the mass of the observed star, $mass_2$ the mass of the spectroscopically detected companion, and $G$ the gravitational constant. In addition,
also the minimum semi-major axis:
\begin{equation}\label{eq:a_orbit}
 a\sin{i} = \frac{K~P~(1-e^{2})^{1/2}}{2\pi}.
\end{equation}
of the orbit of the observed star around the barycenter of the binary system, was derived.

For double-lined spectroscopic binaries (SB2) the minimum semi-major axes of the orbits of both of their components around the common barycenter were calculated using Equation\,\ref{eq:a_orbit}. Furthermore, the minimum mass of these stars:
\begin{equation}
mass_{1/2} \sin^3 i = \frac{(1 - e^2)^3/2 P}{2 \pi G} K_{2/1} (K_{1} + K_{2})^2
\end{equation}
was determined.

\subsection{7\,CrB\,A}\label{sec:7CrBA}

This star is listed in the WDS as the primary component of a wide binary system (see \mbox{Table\,\ref{tab:7CrB_wds}\hspace{-1.5mm}}). The Gaia DR3 astrometry of 7\,CrB\,A and B confirms the companionship of these two stars as they have similar parallaxes (\mbox{$\varpi_{\text{A}}=6.43\pm0.10$\,mas}, and \mbox{$\varpi_{\text{B}}=6.41\pm0.05$\,mas}) and proper motions (\mbox{$\mu^{*}_{\alpha,\text{A}}=-15.51\pm0. 09$\,mas/yr}, \mbox{$\mu_{\delta,\text{A}}=-7.47\pm0.12$\,mas/yr}, \mbox{$\mu^{*}_{\alpha,\text{B}}=-14.01\pm0. 04$\,mas/yr}, \mbox{$\mu_{\delta,\text{B}}=-3.92\pm0.05$\,mas/yr}). 7\,Crb\,B has a cpm-index of about 8 and the differential proper motion of the companion to 7\,CrB\,A is smaller than its estimated escape velocity. Further companions of 7\,CrB\,A are not listed in the Gaia DR3. Thus 7\,CrB\,A and 7\,CrB\,B (see also next subsection) form a binary system with a projected separation of about 990\,au.

\begin{table}[!h]
\caption{Components of 7\,CrB\,A, listed in the WDS, with their angular separation $\rho$ and position angle $\theta$, measured in the first and last observing epoch.}
\centering
\resizebox{\hsize}{!}{\begin{tabular}{ccccccc}
\toprule
Comp. & Date 1 & $\rho_1$ [$''$]  & $\theta_1$ [$\deg$]  & Date 2 & $\rho_2$ [$''$] & $\theta_2$ [$\deg$]  \\ \midrule
B & 1779 & 5.5 & 296 & 2024 & 6.3 & 306  \\
\bottomrule
\end{tabular}}\label{tab:7CrB_wds}
\end{table}

\begin{figure}[h!]
\centering
\resizebox{\hsize}{!}{\includegraphics{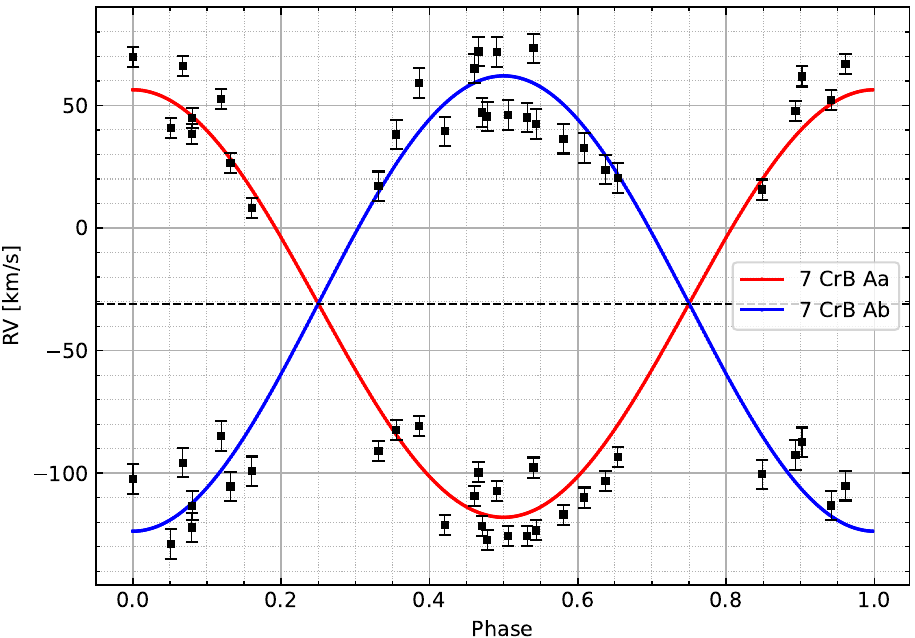}}
\flushright
\resizebox{0.99\hsize}{!}{\includegraphics{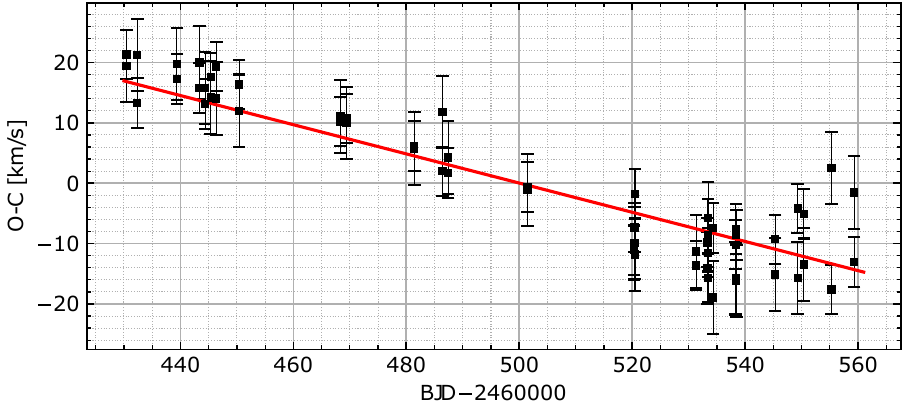}}
\centering
\resizebox{\hsize}{!}{\includegraphics{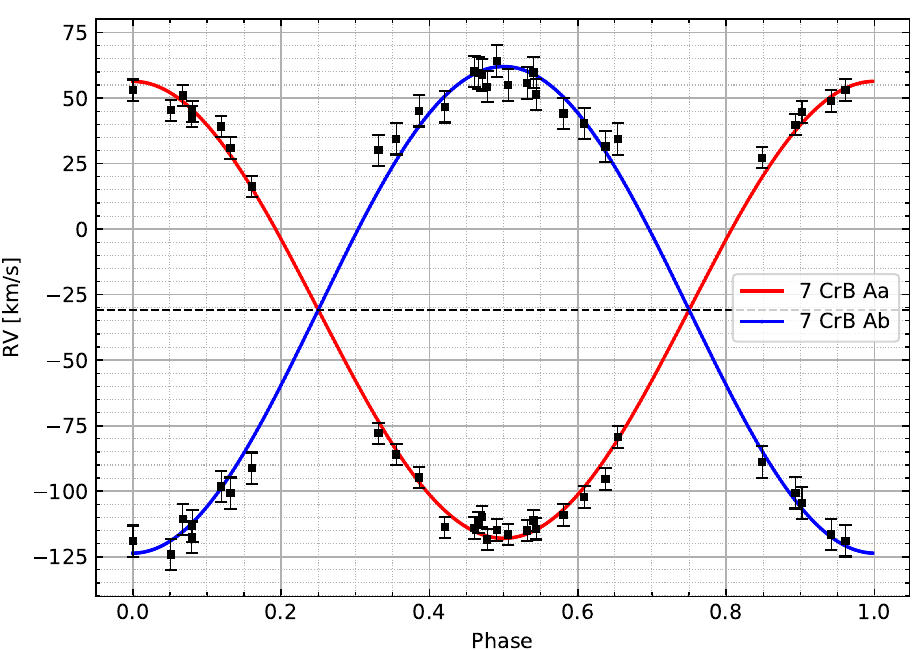}}
\caption{\textbf{Top panel:} Preliminary orbital fit for 7\,CrB\,Aa (red) and 7\,CrB\,Ab (blue), which still shows a large scattering. \textbf{Central panel:} A linear trend (red line; see Equation\,\eqref{eq:trend_7CrBA} for the exact function) was observed in the RV residuals of 7\,CrB\,Aa and Ab. \textbf{Bottom panel:} The trend-corrected RV data and the best-fitting Keplerian orbital solution for 7\,CrB\,Aa (red) and 7\,CrB\,Ab (blue). The systemic velocity is indicated by the dashed black line. The orbital period is \mbox{$P=1.7236\pm0.0002$ days}.}\label{fig:7CrBA_scatter}
\end{figure}

7\,CrB\,A itself is an SB2 system, and a clear splitting of the absorption lines is present in most of the FLECHAS spectra of the star. A total of 30 RV measurements of the two components of the 7\,CrB\,A system were taken over a period of \mbox{129 days}. The individual RV measurements of the two stars are listed in \mbox{Table\,\ref{tab:7CrBA}\hspace{-1.5mm}}. For this target, it proved difficult to find a good Keplerian orbit solution. As shown in the upper panel of \mbox{Figure\,\ref{fig:7CrBA_scatter}\hspace{-1.5mm}}, the RV measurements of both components still scatter strongly around the found best-fitting orbital solution. The orbit fit for 7\,CrB\,Aa has $\chi^{2}_{\text{red}}=6.45$, the one for 7\,CrB\,Ab even $\chi^{2}_{\text{red}}=8.89$, so both are not well-fitting orbital solutions. Therefore, we examined the residuals \mbox{(O--C)} of the determined orbit fits in detail, which are shown in the middle panel of \mbox{Figure\,\ref{fig:7CrBA_scatter}\hspace{-1.5mm}}. The residuals show a significant linear trend in the form:
\begin{align}\label{eq:trend_7CrBA}
\text{RV}_\text{trend} = A + B \times (\text{BJD}-2460000)
\end{align}
with the intercept \mbox{$A=120.9\pm10.2$\,km/s} and the slope \mbox{$B=-0.24\pm0.02$\,(km/s)/day}. This trend in the RV residuals clearly indicates the presence of an additional companion revolving around the spectroscopic binary on a wide (few au) orbit. Thus, 7\,CrB\,A is indeed a hierarchical triple star system.

To accurately determine the orbital elements of 7\,CrB\,Aa and Ab, the determined trend was then subtracted from the original RV data of the two components, using the RV in the middle of the observation period ($\text{BJD}=501.44716$) as a reference. This removed most of the RV scatter, as shown in \mbox{Figure\,\ref{fig:7CrBA_scatter}\hspace{-1.5mm}}. The orbital fits of the trend-corrected RV data have $\chi^{2}_{\text{red}}=1.08$ for 7\,Crb\,Aa, and $\chi^{2}_{\text{red}}=1.21$ for 7\,CrB\,Ab, i.e. the derived Keplerian orbital solutions are now both in good agreement with the given RV measurements.

In the orbit fitting of the trend-corrected RV data, all Keplerian orbital elements were initially left unconstrained as usual, resulting in an orbit solution that does not differ significantly from a circular one (\mbox{$e=0.00^{+0.02}_{-0.00}$}). We therefore set $e=0$ and then use \mbox{$\omega_{\text{Aa}} = 0\,^\circ$} and \mbox{$\omega_{\text{Ab}} = 180\,^\circ$}.

The elements of the final orbital solution for 7\,CrB\,Aa and Ab are summarized in \mbox{Table\,\ref{tab:elements}\hspace{-1.5mm}}. The best-fitting Kepler orbit obtained implies a minimum semi-major axis and mass:\newline \mbox{$a\sin{i} = 2.98\pm0.05~\text{R}_\odot$}, \mbox{$m\sin^{3}{i} = 0. 537\pm0.006~\text{M}_\odot$}\newline for 7\,CrB\,Aa  and\newline \mbox{$a\sin{i} = 3.16\pm0.05~\text{R}_\odot$}, \mbox{$m\sin^{3}{i} = 0.505\pm0.006~\text{M}_\odot$}\newline for 7\,CrB\,Ab, respectively.

\begin{table}[h!]
\caption{For 7\,CrB\,A, 30 RV measurements were carried out over a period of \mbox{129 days}. The measurements scatter strongly around the best-fitting Keplerian orbital solution, due to a superimposed linear trend in the RV, which was subsequently subtracted from the measurements ($\text{RV}_{\text{corr}}$). The typical error of each RV measurement of the Aa component is \mbox{$\Delta\text{RV} = 4.10$\,km/s}, or \mbox{$\Delta\text{RV}=6.00$\,km/s} for the Ab component. The table contains the SNR of all spectra.}
\centering
\resizebox{\hsize}{!}{\begin{tabular}{cccccc}
\toprule
$\text{BJD}-2460000$ & $\text{RV}^{(\text{Aa})}$ & $\text{RV}^{(\text{Aa})}_{\text{corr}}$ & $\text{RV}^{(\text{Ab})}$ & $\text{RV}^{(\text{Ab})}_{\text{corr}}$ &  SNR\\
& [km/s]                    &  [km/s]                                 &  [km/s]                   &  [km/s]                                 &\\
\midrule
430.47210 & $61.82$ & $44.66$ & $-87.32$ & $-104.49$ &  126 \\
432.36450 & $69.70$ & $52.99$ & $-102.38$ & $-119.09$ &  105 \\
439.37463 & $65.95$ & $50.94$ & $-95.71$ & $-110.72$ &  123 \\
443.37099 & $-80.76$ & $-94.80$ & $59.21$ & $45.17$ &  113 \\
444.36213 & $66.92$ & $53.12$ & $-105.11$ & $-118.91$ &  110 \\
445.36038 & $-97.67$ & $-111.23$ & $73.31$ & $59.75$ &  123 \\
446.35807 & $52.51$ & $39.19$ & $-84.85$ & $-98.17$ &  126 \\
450.40350 & $-99.65$ & $-111.99$ & $71.97$ & $59.63$ &  115 \\
468.37669 & $47.84$ & $39.85$ & $-92.62$ & $-100.62$ &  120 \\
469.40618 & $-107.11$ & $-114.86$ & $71.87$ & $64.12$ &  122 \\
481.41943 & $-109.16$ & $-114.00$ & $65.02$ & $60.18$ &  142 \\
486.40871 & $-82.40$ & $-86.04$ & $38.10$ & $34.47$ &  134 \\
487.41902 & $52.26$ & $48.87$ & $-113.20$ & $-116.59$ &  108\\
501.44716 & $44.92$ & $44.92$ & $-113.23$ & $-113.23$ &  108 \\
520.35736 & $40.77$ & $45.34$ & $-128.79$ & $-124.22$ &  114 \\
520.40604 & $38.38$ & $42.97$ & $-122.14$ & $-117.56$ &  104 \\
520.49585 & $26.41$ & $31.01$ & $-105.38$ & $-100.77$ &  138 \\
531.33562 & $-121.02$ & $-113.79$ & $39.46$ & $46.69$ &  108 \\
533.33468 & $-116.86$ & $-109.15$ & $36.40$ & $44.11$ &  146 \\
533.38321 & $-109.92$ & $-102.20$ & $32.60$ & $40.32$  & 138 \\
533.43252 & $-103.15$ & $-95.41$ & $23.80$ & $31.53$ & 123 \\
534.33443 & $8.26$ & $16.21$ & $-99.18$ & $-91.23$ &  119 \\
538.32875 & $-127.33$ & $-118.41$ & $45.43$ & $54.35$ &  122 \\
538.37742 & $-125.50$ & $-116.57$ & $46.08$ & $55.01$ &  109 \\
538.44230 & $-123.29$ & $-114.34$ & $42.39$ & $51.34$ &  109 \\
545.31564 & $-125.56$ & $-114.95$ & $45.10$ & $55.71$ &  104 \\
549.30932 & $15.69$ & $27.26$ & $-100.43$ & $-88.86$ &  96 \\
550.38204 & $-121.59$ & $-109.76$ & $47.06$ & $58.89$ &  59 \\
555.31172 & $-90.93$ & $-77.91$ & $17.06$ & $30.08$ &  130 \\
559.31557 & $-93.26$ & $-79.27$ & $20.35$ & $34.34$ &  110 \\
\bottomrule
\end{tabular}}\label{tab:7CrBA}
\end{table}

\subsection{7\,CrB\,B}

As described in the previous section, 7\,CrB\,B is a wide companion of the spectroscopically detected hierarchical triple star system 7\,CrB\,A.

The RV of 7\,CrB\,B was monitored with FLECHAS in 25 observation epochs within a period of \mbox{125 days}. The individual RV measurements of the star are listed in \mbox{Table\,\ref{tab:7CrBB}\hspace{-1.5mm}}. As in the case of 7\,CrB\,A, also 7\,CrB\,B shows a significant RV variation over time.

\begin{table}[h!]
\caption{For 7\,CrB\,B, 25 RV measurements were carried out over a period of \mbox{125 days}. The typical error of each RV measurement is \mbox{$\Delta\text{RV} = 2.70$\,km/s}. The table includes the SNR of all spectra.}
\centering
\begin{tabular}{ccc}
\toprule
$\text{BJD}-2460000$ & RV [km/s]  & SNR \\ \midrule
430.46621 & $-16.78$ &  103\\
432.37206 & $-21.08$ &  151\\
439.38085 & $-17.33$ &  165\\
443.37700 & $-16.27$ &  139\\
444.36890 & $-25.59$ &  90\\
445.36646 & $-23.91$ &  152\\
446.36633 & $-13.73$ &  128\\
450.40960 & $-12.82$ &  150\\
468.39258 & $-22.07$ &  158\\
469.41232 & $-21.83$ &  169\\
473.43221 & $-27.21$ &  150\\
486.41554 & $-21.11$ &  192\\
487.42517 & $-10.88$ &  170\\
501.45444 & $-22.11$ &  159\\
516.43550 & $-14.94$ &  125\\
520.36338 & $-9.49$ &  170\\
531.34189 & $-15.00$ &  142\\
533.34071 & $-16.88$ &  209\\
535.34299 & $-13.02$ &  187\\
538.33470 & $-28.21$ &  196\\
545.32218 & $-19.40$ &  150\\
547.39155 & $-19.91$ &  141\\
549.31554 & $-18.57$ &  142\\
550.38982 & $-20.96$ &  145\\
555.31787 & $-19.13$ &  177\\
\bottomrule
\end{tabular}\label{tab:7CrBB}
\end{table}

\begin{figure}[h!]
\centering
\includegraphics[width=1\linewidth]{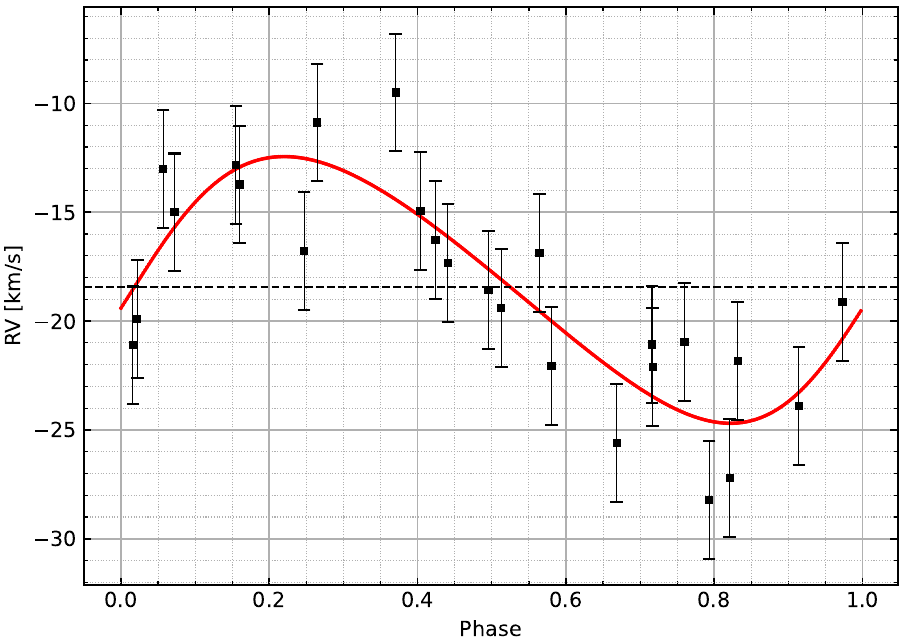}
\caption{The best-fitting Keplerian orbital solution (red line) of the FLECHAS RV data of 7\,CrB\,B with a period of \mbox{$P=4.06\pm0.01$ days}. The systemic velocity is indicated by the dashed black line.}\label{fig:7CrBB}
\end{figure}

A Keplerian orbital solution could be fitted to the FLECHAS RV data of 7\,CrB\,B, which agrees well with the RV measurements (\mbox{$\chi^{2}_{\text{red}}=1.03$}). The elements of this solution are summarized in Table\,\ref{tab:elements}. The solution presented here implies a minimum semi-major axis of the orbit of the primary component around the barycenter of the system \mbox{$a\sin{i} = 0.49\pm0.10~\text{R}_\odot$}, and a mass function of the system \mbox{$f(m) = (9.31\pm5.53) \times 10^{-5}~\text{M}_\odot$}.

Hence, 7\,CrB is a quintuple star system composed of the spectroscopically identified triple star system 7\,CrB\,A (see previous section), which is orbited by the SB1 system 7\,CrB\,B on a wide orbit with a projected separation of about 990\,au.

\subsection{HD\,214007}

\begin{figure}[h!]
\centering
\includegraphics[width=1\linewidth]{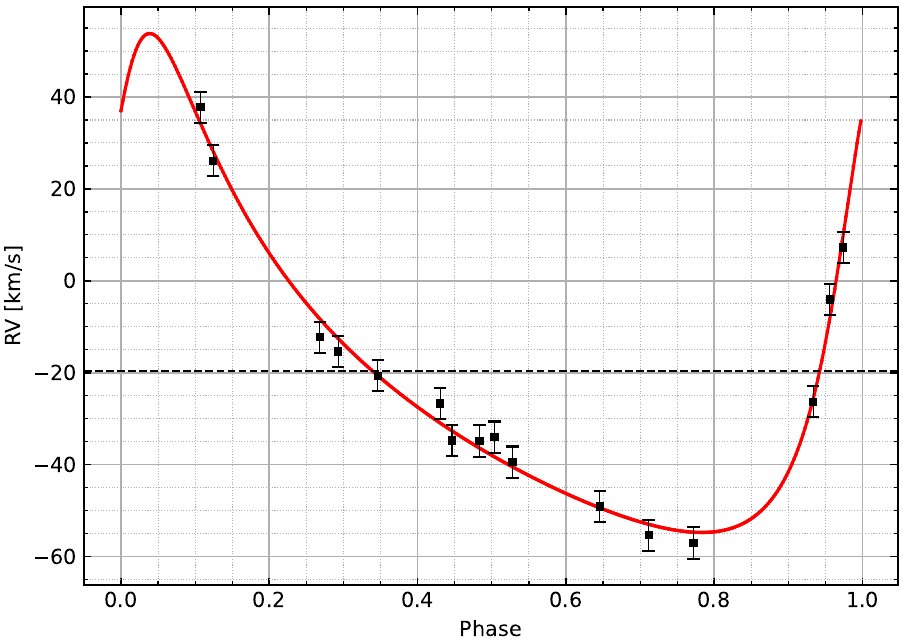}
\caption{Best-fitting Keplerian orbital solution (red line) on the FLECHAS RV data of HD\,214007, which has a period of \mbox{$P=10.16\pm0.01$ days}. The systemic velocity is indicated by the dashed black line.}\label{fig:HD214007}
\end{figure}

HD\,214007 is not listed in the WDS and no companions of the star can be found in the Gaia DR3. The RV of the star was measured 16 times with FLECHAS during an observation time of \mbox{130 days}. The individual RV measurements are listed in \mbox{Table\,\ref{tab:HD214007}\hspace{-1.5mm}} and show a considerable variation. Therefore, we have tried to fit a Keplerian orbital solution to the RV data. The best Kepler fit is shown in \mbox{Figure\,\ref{fig:HD214007}\hspace{-1.5mm}} and agrees well (\mbox{$\chi^{2}_{\text{red}}=1.05$}) with the RV data. The Keplerian elements of this solution are listed in \mbox{Table\,\ref{tab:elements}\hspace{-1.5mm}}. According to this fit, the orbit of the primary component around the barycenter of the system has a minimum semi-major axis \mbox{$a\sin{i} = 9.39\pm0.67\,\text{R}_\odot$} and the mass function of the system is \mbox{$f(m) = 0.11\pm0.02\,\text{M}_\odot$}.

\begin{table}[h!]
\caption{For HD\,214007, 16 RV measurements were carried out over a period of \mbox{130 days}. The typical error of each RV measurement is \mbox{$\Delta\text{RV} = 3.40$\,km/s}. The table includes the SNR of all spectra.}
\centering
\begin{tabular}{ccc}
\toprule
$\text{BJD}-2460000$ & RV [km/s]  & SNR \\ \midrule
430.51326  & $-49.13$ &  94\\
439.48140  & $-39.46$ &  124 \\
445.54396  & $26.11$  &  117 \\
451.51301  & $-55.39$ &  104 \\
469.52137  & $-34.88$ &  112 \\
516.52393  & $37.77$  &  119 \\
520.55057  & $-34.02$ &  151 \\
528.57551  & $-15.36$ &  121\\
533.44334  & $-57.06$ &  171 \\
535.49831  & $7.27$ &  130\\
538.48836  & $-12.30$ &  155 \\
545.47773  & $-4.09$ &  135\\
549.44824  & $-20.65$ & 143\\
550.46464  & $-34.76$ &  167\\
555.41571  & $-26.32$ &  144 \\
560.46614  & $-26.70$ &  110\\
\bottomrule
\end{tabular}\label{tab:HD214007}
\end{table}

\subsection{$\iota$\,Her}

The WDS lists two companions of this star (see \mbox{Table\,\ref{tab:iotHer_wds}\hspace{-1.5mm}}), neither of which can be confirmed with astrometric data from the Gaia DR3. The Ab component, whose angular distance to $\iota$\,Her corresponds to a projected separation of about 30\,au, is too close to the star to be listed in the Gaia DR3. Since only one observational epoch of this component is given in the WDS, its companion status remains inconclusive. The proposed B component is detected in the Gaia DR3, but only has a cpm-index of about 0.9 and a parallax (\mbox{$\varpi=1.22\pm0.01$\,mas}) that differs significantly from that of $\iota$\,Her (\mbox{$\varpi=6.52\pm0.20$\,mas}). Therefore, this component can be clearly ruled out as a true companion of the star. Further companions of $\iota$\,Her are not listed in the Gaia DR3.

\begin{table}[h!]
\caption{Components of $\iota$\,Her, listed in the WDS, with their angular separation $\rho$ and position angle $\theta$, measured in the first and last observing epoch.}
\centering
\resizebox{\hsize}{!}{\begin{tabular}{ccccccc}
\toprule
Comp. & Date 1 & $\rho_1$ [$''$]  & $\theta_1$ [$\deg$] & Date 2 &  $\rho_2$ [$''$] & $\theta_2$ [$\deg$]  \\ \midrule
Ab & 1975 & 0.2 & 6  & -- & -- &-- \\
B &  1893 & 115.1 & 49 & 2023 & 117.2 & 48  \\
\bottomrule
\end{tabular}}\label{tab:iotHer_wds}
\end{table}

\begin{figure}[h!]
\centering
\includegraphics[width=1\linewidth]{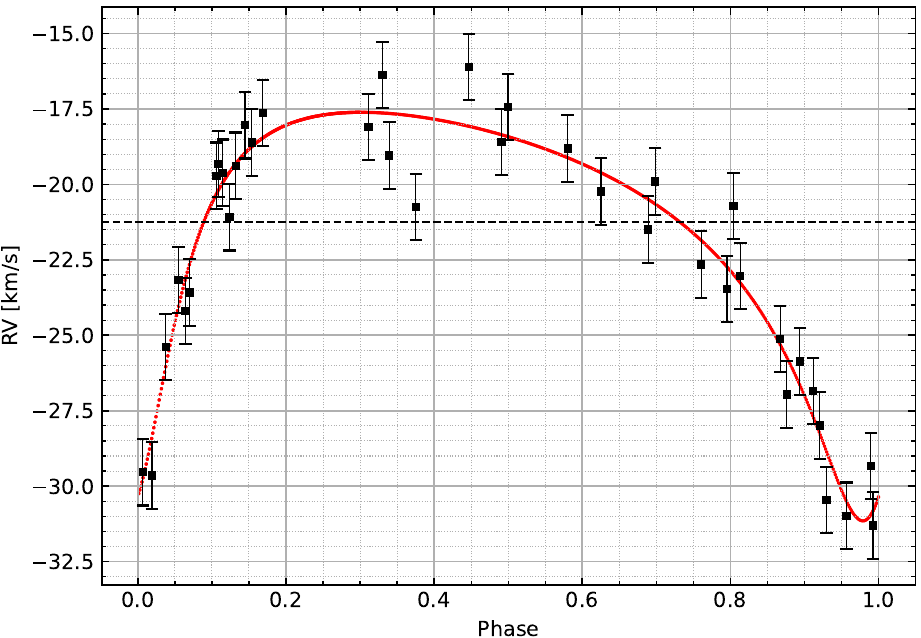}
\caption[$\iota$ Her: orbit solution]{Fitted Keplerian orbital solution for $\iota$\,Her, which has a period of \mbox{$P=111.5\pm0.1$ days}. The systemic velocity is indicated by the dashed black line.}\label{fig:iotHer}
\end{figure}

For $\iota$ Her, a total of 38 RV measurements were carried out with FLECHAS over a period of \mbox{130 days}. They are all summarized in \mbox{Table\,\ref{tab:iotHer}\hspace{-1.5mm}}. The Keplerian orbit solution fitted to these RV data points is shown in \mbox{Figure\,\ref{fig:iotHer}\hspace{-1.5mm}} and the corresponding orbital elements are listed in \mbox{Table\,\ref{tab:elements}\hspace{-1.5mm}}. With these elements we derived the minimum semi-major axis of the orbit of the primary componet of the system \mbox{$a\sin{i} = 12.68\pm0.74\,\text{R}_\odot$} as well as the mass function of the system \mbox{$f(m) = (2.20\pm0.43) \times 10^{-3}\,\text{M}_\odot$}.

\begin{table}[h!]
\caption{For $\iota$\,Her, 38 RV measurements were carried out over a period of \mbox{130 days}. The typical error of each RV measurement is \mbox{$\Delta\text{RV} = 1.10$\,km/s}. The table includes the SNR of all spectra.}
\centering
\begin{tabular}{ccc}
\toprule
$\text{BJD}-2460000$ & RV [km/s]  & SNR  \\ \midrule
430.49612  & $-29.33$ &  207\\
432.42794  & $-29.54$ &  196\\
439.46529  & $-23.58$ &  225 \\
443.47412  & $-19.71$ &  176\\
444.45129  & $-19.61$ &  197 \\
445.43843  & $-21.09$ &  197 \\
446.39566  & $-19.38$ &  147\\
450.48479  & $-17.64$ &  199 \\
466.37568  & $-18.09$ &  227 \\
468.49466  & $-16.37$ &  212 \\
469.51033  & $-19.04$ &  206\\
473.52672  & $-20.74$ &  167\\
481.50329  & $-16.10$ &  298 \\
486.38542  & $-18.59$ &  216 \\
487.38395  & $-17.44$ &  239 \\
496.37381  & $-18.81$ &  194 \\
501.38235  & $-20.23$ &  236 \\
508.48438  & $-21.49$ &  297 \\
509.52354  & $-19.90$ &  206 \\
516.44620  & $-22.65$ &  161 \\
520.35464  & $-23.46$ &  200 \\
521.35658  & $-20.71$ &  216 \\
522.35895  & $-23.03$ &  259 \\
528.35530  & $-25.13$ &  171 \\
529.34059  & $-26.96$ &  216 \\
531.33324  & $-25.87$ &  175 \\
533.33090  & $-26.85$ &  269 \\
534.33147  & $-27.99$ &  218 \\
535.33285  & $-30.46$ &  201 \\
538.32610  & $-30.98$ &  241 \\
542.33106  & $-31.30$ &  155\\
545.31240  & $-29.65$ &  215 \\
547.37931  & $-25.38$ &  202 \\
549.30699  & $-23.16$ &  228 \\
550.32532  & $-24.20$ &  249 \\
555.30984  & $-19.33$ &  238 \\
559.31321  & $-18.04$ &  219 \\
560.35906  & $-18.61$ &  189 \\
\bottomrule
\end{tabular}\label{tab:iotHer}
\vspace{1.5cm}
\end{table}

\subsection{HD\,201433\,A}\label{sec:HD201433}

Three potential companions are listed in the WDS for HD\,201433, which are summarized in \mbox{Table\,\ref{tab:HD201433_wds}\hspace{-1.5mm}}.

\begin{table}[h!]
\caption{Components of HD\,201433, listed in the WDS, with their angular separation $\rho$ and position angle $\theta$, as measured in the first and last observing epoch.}
\centering
\resizebox{\hsize}{!}{\begin{tabular}{ccccccc}
\toprule
Comp. & Date 1 & $\rho_1$ [$''$]  & $\theta_1$ [$\deg$]  & Date 2 &  $\rho_2$ [$''$] & $\theta_2$ [$\deg$]  \\ \midrule
B & 1783 & 4.0 & 315 & 2020 & 3.3 & 303  \\
C & 1865 & 57.7& 225 & 2021 & 59.0 & 229  \\
D & 1894 & 74.2 & 69 & 2002 & 73.1 & 66  \\
\bottomrule
\end{tabular}}\label{tab:HD201433_wds}
\end{table}

However, only the B component, which has a projected separation of about 410\,au to HD\,201433, can be confirmed as a true companion of the star, as its Gaia DR3 parallax and proper motion (\mbox{$\varpi=8.45\pm0.03$\,mas}, \mbox{$\mu^{*}_{\alpha,\text{B}}=24. 00\pm0.02$\,mas/yr}, \mbox{$\mu_{\delta,\text{B}}=-25.16\pm0.03$\,mas/yr}) are well comparable to those of HD\,201433\,A (\mbox{$\varpi=8.20\pm0.10$\,mas}, \mbox{$\mu^{*}_{\alpha}=24.65\pm0.08$\,mas/yr}, and \mbox{$\mu_{\delta}=-20.65\pm0.10$\,mas/yr}. HD\,201433\,B has a cpm-index of about 15, and the differential proper motion of the companion to HD\,201433\,A is smaller than its estimated escape velocity. Components C and D, on the other hand, can both be excluded as companions of HD\,201433\,A, as they have a cpm-index of only 0.9 and 1.4, respectively. Furthermore, their parallax (\mbox{$\varpi_\text{C}=1.27\pm0.01$\,mas}, \mbox{$\varpi_\text{D}=1.75\pm0.03$\,mas}) differs significantly from that of the star. Further companions of HD\,201433\,A are not listed in the Gaia DR3.

\begin{figure}[h!]
\centering
\includegraphics[width=1\linewidth]{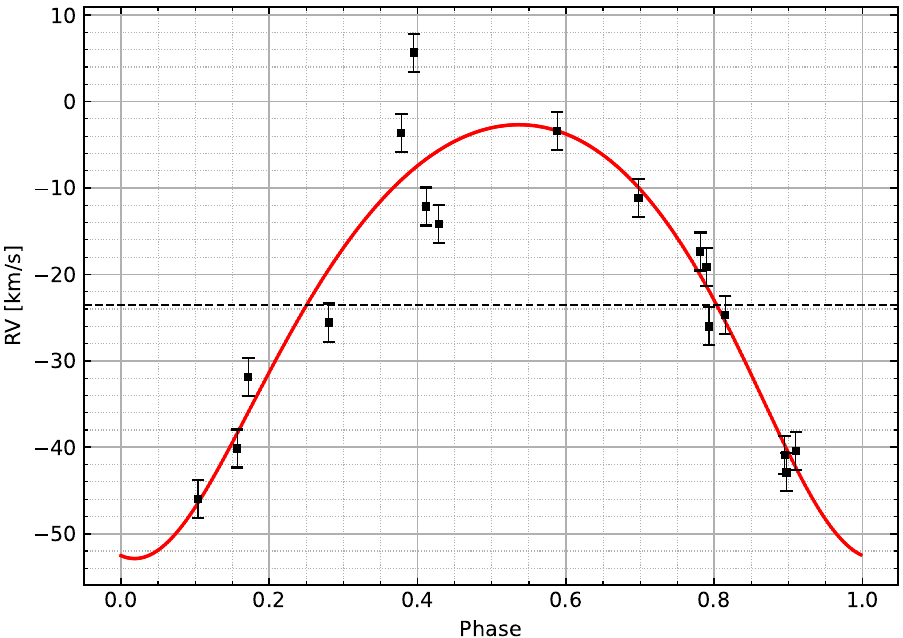}
\caption{Preliminary Keplerian orbit fitted to all RV measurements of HD\,201433\,A. The derived orbital solution has period of \mbox{$P=3.318\pm0.004$ days}. The systemic velocity is indicated by the dashed black line.}\label{fig:HD201433_all}
\end{figure}

For HD\,201433\,A, a total of 17 RV measurements were performed with FLECHAS over a period of \mbox{115 days}, which are summarized in \mbox{Table\,\ref{tab:HD201433}\hspace{-1.5mm}}. We have fitted a Keplerian orbit solution to all these data, which is shown in \mbox{Figure\,\ref{fig:HD201433_all}\hspace{-1.5mm}}.

\begin{table}[h!]
\caption{For HD\,201433\,A, 17 RV measurements were carried out over a period of \mbox{115 days}. Measurements that deviate significantly from the orbital solution, considering all RV data, which is shown in \mbox{Figure\,\ref{fig:HD201433_all}\hspace{-1.5mm}}, are written in gray. The typical error of each RV measurement is \mbox{$\Delta\text{RV}=2.20$\,km/s}. The table also contains the SNR of all spectra.}
\centering
\begin{tabular}{ccc}
\toprule
$\text{BJD}-2460000$ & RV [km/s] &  SNR\\ \midrule
445.50990  & $-19.13$ &   94\\
451.47733  & $-3.38$ &   100\\
486.54315  & $-40.14$ &   151\\
\textcolor{gray}{500.54992}  & \textcolor{gray}{$-3.65$} &   \textcolor{gray}{111}\\
508.52379  & $-17.35$ &  109\\
\textcolor{gray}{516.45563}  & \textcolor{gray}{$-31.87$} &   \textcolor{gray}{102}\\
\textcolor{gray}{520.51409}  & \textcolor{gray}{$5.67$} &  \textcolor{gray}{145}\\
528.54257  & $-24.72$ &   108\\
531.47316  & $-11.15$ &   153\\
533.40637  & $-25.57$ &   161\\
535.45312  & $-42.90$ &   149\\
538.42472  & $-26.02$ &   154\\
545.39957  & $-40.91$ &  132\\
549.41059  & $-46.00$ &   108\\
550.43157  & $-12.14$ &   148\\
555.40373  & $-40.42$ &   130\\
\textcolor{gray}{560.44223}  & \textcolor{gray}{$-14.16$} &  \textcolor{gray}{111}\\
\bottomrule
\end{tabular}\label{tab:HD201433}
\end{table}

As can be seen in \mbox{Figure\,\ref{fig:HD201433_all}\hspace{-1.5mm}}, some RV measurements scatter significantly around the orbital solution presented (\mbox{$\chi^{2}_\text{red}=7.78$}). Therefore, as already done for 7\,CrB\,A (see Section\,\ref{sec:7CrBA}), we subtracted the derived Keplerian orbital solution from the RV measurements to examine the residuals for a possible linear trend that could explain the large deviation of some RV measurements from the orbit fit. However, no significant trend could be detected in the RV residuals. So there must be another reason for the RV scattering. In this context, it is worth mentioning that the large RV scattering of the star is also noted in the 9th Catalog of Spectroscopic Double Star Orbits \citep[SB9 from here on, ][]{pourbaix2004}, where it is suggested that both components of HD\,201433\,A could be pulsating variables.

\begin{figure}[h!]
\centering
\includegraphics[width=1\linewidth]{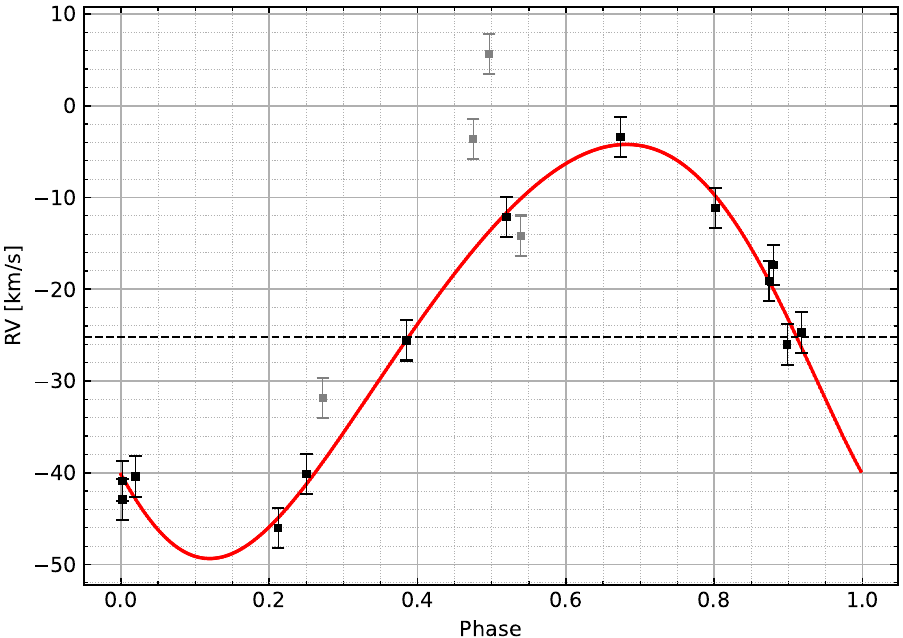}
\caption{Keplerian orbit solution fitted on all RV data of HD\,201433\,A, with identified outliers (marked in gray) excluded from the fit. This orbital solution has a period of \mbox{$P=3.325^{+0.001}_{-0.002}$ days}.}\label{fig:HD201433}
\end{figure}

Assuming that this is true, we attempted to find a better-fitting orbital solution by excluding from the fit all RV measurements that deviate by more than $3\sigma$ from the orbital fit based on all RV measurements. The derived elements of both orbital solutions, which do not differ significantly from each other, are listed in \mbox{Table\,\ref{tab:elements}\hspace{-1.5mm}}. The resulting final orbital solution ($\chi^{2}_\text{red}=1.05$) is shown in (\mbox{Figure\,\ref{fig:HD201433}\hspace{-1.5mm}}). From this orbital fit, we obtained \mbox{$a\sin{i}=1.47\pm0.08\,\text{R}_\odot$} for the minimum semi-major axis of the primary of the binary system and \mbox{$f(m)=(3.86\pm0.61) \times 10^{-4}\,\text{M}_\odot$} for its mass-function. Thus, HD\,201433 is at least a hierarchical triple star system consisting of the SB1 system HD\,201433\,A, which is orbited by the distant companion HD\,201433\,B.

\subsection{Summary}

For all stars with significant RV variation identified in our RV monitoring program, the elements and the $\chi^{2}_{\text{red}}$-value of the best-fitting Kepler orbital solution are summarized in \mbox{Table\,\ref{tab:elements}\hspace{-1.5mm}}.

\begin{table*}[t!]
\caption{Elements and $\chi^{2}_{\text{red}}$-value of the best-fitting Kepler orbital solutions determined for all targets that showed significant variation in their RV during our FLECHAS spectroscopy monitoring campaign. In the case of HD\,201433\,A, two solutions are given. The one written in black considers all RV data, the one written in gray excludes some outliers (see Section\,\ref{sec:HD201433} for more details).}
\centering
\resizebox{\hsize}{!}{\begin{tabular}{cccccccc}
\toprule
Target & $P$ [d] & $T$ [$\text{BJD}-2460000$] & $e$ &  $\omega$ [$\deg$] &  $K$ [km/s] & $\gamma$ [km/s] & $\chi^{2}_{\text{red}}$\\ \midrule
7\,CrB\,Aa & $1.7236^{+0.0002}_{-0.0002}$ & $428.92^{+0.01}_{-0.01}$ & 0 (fixed) & 0 (fixed) & $87.19^{+1.10}_{-1.09}$ & $-30.79^{+0.66}_{-0.66}$ & 1.08\\
7\,CrB\,Ab & $1.7236^{+0.0002}_{-0.0002}$ & $428.92^{+0.01}_{-0.01}$ & 0 (fixed) & 180 (fixed) & $92.86^{+1.09}_{-1.09}$ & $-30.79^{+0.66}_{-0.66}$ & 1.21\\
7\,CrB\,B & $4.06^{+0.01}_{-0.01}$ & $429.46^{+0.14}_{-0.14}$ & $0.16^{+0.14}_{-0.16}$ & $262.2^{+66.1}_{-55.1}$ & $6.1^{+0.8}_{-0.8}$ & $-18.43^{+0.50}_{-0.51}$ & 1.03\\
HD\,214007   &  $10.16^{+0.01}_{-0.01}$ & $423.95^{+0.11}_{-0.11}$ & $0.51^{+0.03}_{-0.03}$ & $313.7^{+4.4}_{-4.7}$& $54.3^{+3.8}_{-2.8}$ & $-19.62^{+1.22}_{-1.23}$ & 1.05\\
$\iota$\,Her & $111.5^{+0.1}_{-0.1}$ & $320.13^{+1.33}_{-1.33}$ & $0.53^{+0.03}_{-0.03}$ & $208.3^{+5.2}_{-5.0}$& $6.8^{+0.4}_{-0.3}$ & $-21.23^{+0.20}_{-0.20}$ & 1.15\\
HD\,201433\,A & $3.318^{+0.004}_{-0.004}$ & $442.89^{+0.29}_{-0.29}$ & $0.17^{+0.11}_{-0.10}$ &  $170.5^{+34.5}_{-39.2}$ & $25.1^{+3.2}_{-2.7}$ & $-23.57^{+1.47}_{-1.66}$ & 7.78\\
\textcolor{gray}{HD\,201433\,A} & \textcolor{gray}{$3.325^{+0.001}_{-0.002}$} & \textcolor{gray}{$442.61^{+0.16}_{-0.16}$} & \textcolor{gray}{$0.12^{+0.03}_{-0.03}$} &  \textcolor{gray}{$126.4^{+17.6}_{-20.6}$} & \textcolor{gray}{$22.6^{+1.0}_{-0.9}$} & \textcolor{gray}{$-25.16^{+0.57}_{-0.58}$} & \textcolor{gray}{1.05}\\
\bottomrule
\end{tabular}}\label{tab:elements}
\end{table*}

\section{Conclusions}\label{sec:orb_elem}

For this study, we used FLECHAS at the University Observatory Jena to observe 13 selected stars and accurately determine their mean RV.

The RV of 17\,Dra\,A (\mbox{$\text{RV}=-14.81\pm1.20$\,km/s}) remained constant during our 79-day RV observing program of the star. Therefore, very close companions of 17\,Dra\,A can be excluded. However, according to information from the WDS and Gaia DR3, 17\,Dra is actually a hierarchical quadruple system consisting of two binary stars separated by about 11000\,au, with projected separations of 390 and 1500\,au, respectively.

\begin{figure}[h!]
\centering
\includegraphics[width=1\linewidth]{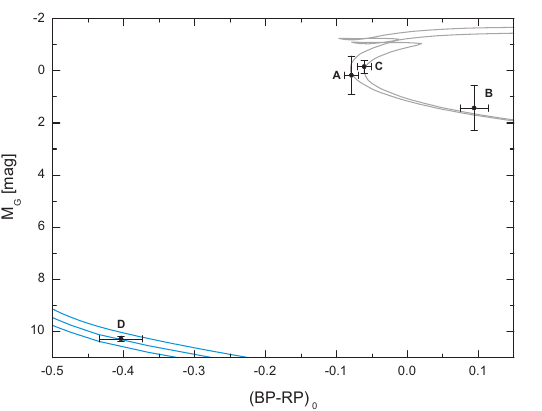}
\caption{Color-magnitude diagram of all components of the 17\,Dra system. The isochrones of stars with ages of 270 (left) and 310\,Myr (right) are shown as gray lines, and the blue curves are the mass tracks of DA white dwarfs with masses of 0.6 (upper), 0.7 (middle) and 0.8\,$\text{M}_{\odot}$ (lower). The photometry of 17\,Dra\,A and its companions 17\,Dra\,B and C agrees well with that expected for main-sequence stars. In contrast, the photometry of 17\,Dra\,D corresponds to that of DA white dwarfs.}\label{fig:17Dra_Photo}
\end{figure}

In order to determine the nature of the detected companions we derived the absolute magnitude and intrinsic color for all components of the 17\,Dra system using their Gaia DR3 photometry as well as the reddening and parallax of 17\,Dra\,A. We combined our results with those listed for the components in the VizieR database. We plotted all components of the 17\,Dra system in a $(BP-RP)_0$-$M_\text{G}$ color-magnitude diagram, which is shown in \mbox{Figure\,\ref{fig:17Dra_Photo}\hspace{-1.5mm}}. In this diagram, we also show as gray lines isochrones for stars of solar metallicity and ages of 270 and 310\,Myr from the PARSEC 1.2S models\footnote{The models are available online at: \url{https://stev.oapd.inaf.it/PARSEC/}} \citep{bressan2012, chen2014, chen2015, marigo2017, pastorelli2019, pastorelli2020, tang2014}, as well as tracks of DA white dwarfs (blue lines) with masses of 0.6, 0.7, and 0.8\,$\text{M}_\odot$ from the Bergeron et al. evolutionary white dwarf models\footnote{The models are available online at: \url{https://www.astro.umontreal.ca/~bergeron/CoolingModels/}} \citep[version 2021.01.13, ][]{bedard2020, bergeron2011,  blouin2018, holberg2006, kowalski2006, tremblay2011}. While the photometry of 17\,Dra\,A, B, and C agrees well with that expected for main-sequence stars, 17\,Dra\,D is significantly fainter than main-sequence stars but appears much bluer than these stars. The photometry of this star corresponds to that of DA white dwarfs. We therefore conclude that 17\,Dra is an evolved hierarchical quadruple star system with 17\,Dra\,D, which is the former primary component of this system. 17\,Dra\,D is the only white dwarf companion detected in the Gaia DR3 among the targets whose multiplicity is studied in this work. The remaining companions detected directly by Gaia around the targets are main-sequence stars, except for 57\,CnC\,B, which, like its primary, is a giant star.

The RV of HD\,148374 (\mbox{$\text{RV}=-23.52\pm0.23$\,km/s}) remained stable throughout our 79-day spectroscopic monitoring period. Using data from the WDS and Gaia DR3, we classified the star as a binary system with a projected separation of approximately 140\,au.

HD\,169487 is listed in the WDS as a single star, but we identified a wide companion to the star in the Gaia DR3, with a projected separation of approximately 50200\,au. Therefore, HD\,169487 is the primary star of a wide binary system. During our 91-day spectroscopic monitoring of the star, its RV remained constant (\mbox{$\text{RV} = 3.93\pm0.39$\,km/s}).

Using Gaia DR3 astrometry, we proved that 57\,Cnc is a binary system with a projected distance of 180\,au. The RV of the star remained constant during our spectroscopic observations (\mbox{$\text{RV} = -57.68\pm0.28$\,km/s}).

$\gamma$\,And, which has a triple-star companion at a projected separation of approximately 1000\,au, also exhibited a constant RV during our spectroscopic monitoring campaign (\mbox{$\text{RV} = -57.68\pm0.28$\,km/s}).

HD\,11031 is a quintuple star system. Its primary component is a binary system with an orbital period of approximately \mbox{37 years}. There are two distant companions: HD\,11031\,B at a projected separation of approximately 270\,au, which is a spectroscopic binary system itself, and HD\,11031\,C at a separation of about 2800\,au. During our spectroscopic observation campaign, we investigated the RV of the primary component of this system, which exhibited a constant RV of (\mbox{$\text{RV} = -4.48\pm0.59$\,km/s}).

$\kappa$\,And has a substellar companion located at a projected separation of about 50\,au. No other companions of the star are listed in the Gaia DR3, and its RV remained constant (\mbox{$\text{RV} = -18.96\pm1.86$\,km/s}) during our spectroscopic study of the star.

$\lambda$\,Cas is a binary star system for which an orbital solution with a period of approximately \mbox{246 years} was determined. The RV of the star remained constant during our spectroscopic monitoring program (\mbox{$\text{RV} = -19.18\pm2.92$\,km/s}).

7\,Cr\,B is a hierarchical quintuple system. The two stars 7\,CrB\,A and 7\,CrB\,B have a projected separation from each other of approximately 990\,au and were both observed with FLECHAS as part of our RV monitoring program. They both exhibit significant RV variations. 7\,CrB\,A is an SB2 system with an orbital period of about \mbox{1.7 days}. It also shows a linear trend in its periodic RV variation, indicating the presence of an additional companion in an orbit of a few au. In contrast, 7\,CrB\,B is an SB1 system with an orbital period of about \mbox{four days}.

No companions are known for HD\,214007, neither in the WDS nor in the Gaia DR3. During our RV monitoring program of the star, we found that HD\,214007 is an eccentric ($e\sim0.5$) SB1 system with an orbital period of about \mbox{10 days}.

Our spectroscopic observations of the star $\iota$\,Her revealed that its RV varies periodically. The best-fitting orbital solution for the star's RV data shows that it is an eccentric ($e\sim0.5$) SB1 system with an orbital period of approximately \mbox{112 days}. This star together with HD\,214007 are the most eccentric binary systems identified in our spectroscopic monitoring program.

HD\,201433\,A has a wide companion located at a projected separation of about 410\,au, which is listed in the Gaia DR3. The star also has a variable RV, but it was difficult to find an orbital solution that fits the RV data of the star well. The best-fitting orbital solution, which considers all RV measurements, has $\chi^2_\text{red}=7.78$, so it is not of acceptable quality. Since no linear trend in the RV residuals was observed as in 7\,CrB\,A, the RV scatter of HD\,201433\,A was considered intrinsic. The conspicuous RV scattering of the star is also present in the SB9, where it is suggested that the star is a binary consisting of two pulsating variables. Assuming this to be the case, we removed the identified RV outliers to determine the system's best-fitting Keplerian orbital solution. It is worth noting in this context that the elements of the two determined solutions do not differ significantly. According to our orbital solutions, the star is an SB1 system with an orbital period of about \mbox{3.3 days}, making HD\,201433 a hierarchical triple star system at least.

For all stars with variable RV, we provide the mean and standard deviation of their RV measurements in \mbox{Table\,\ref{tab:Kepler_vs_mean}\hspace{-1.5mm}}. The RV standard deviation is at least one order of magnitude larger than the uncertainty of the systemic velocity resulting from Keplerian orbital fitting of these binary systems. This demonstrates that a sufficient number of RV measurements well covering the orbital phase of these binary systems is required to accurately determine their mean RV. This is crucial for precisely deriving the three-dimensional space velocity of these systems, which is necessary for accurately calculating their trajectory in the Milky Way.

\begin{table}[h!]
\caption{The mean and standard deviation of the RV measurements of all targets with significantly variable RV, compared to the determined systemic velocity $\gamma$ of these binary systems. In the case of HD\,201433\,A, the results written in gray exclude some outliers, see Section\,\ref{sec:HD201433} for details.}
\centering
\begin{tabular}{ccc}
\toprule
Target & $\overline{\text{RV}}$ [km/s] & $\gamma$ [km/s] \\ \midrule
7\,CrB\,Aa & $-41.46\pm79.17$ & $-30.79^{+0.66}_{-0.66}$\\
7\,CrB\,Ab & $-19.17\pm76.92$ & $-30.79^{+0.66}_{-0.66}$\\
7\,CrB\,B & $-18.73\pm4.87$ & $-18.43^{+0.50}_{-0.51}$ \\
HD\,214007 & $-21.19\pm27.19$ & $-19.62^{+1.22}_{-1.23}$ \\
$\iota$\,Her & $-22.64\pm4.47$ & $-21.23^{+0.20}_{-0.20}$\\
HD\,201433\,A  & $-23.17\pm15.62$ & $-23.57^{+1.47}_{-1.66}$ \\
\textcolor{gray}{HD\,201433\,A}  & \textcolor{gray}{$-26.91\pm14.00$} & \textcolor{gray}{$-25.16^{+0.57}_{-0.58}$} \\
\bottomrule
\end{tabular}\label{tab:Kepler_vs_mean}
\end{table}

All data recorded as part of our spectroscopic monitoring project is available from us upon request.

\section*{Acknowledgments}
We thank the observers at the University Observatory Jena who have contributed to some of the FLECHAS observations, in particular: L. Cortese, S. H\"{u}ttel, G. Payl\i, and S. Emmerichs. This work has made use of data from the European Space Agency (ESA) mission \textit{Gaia} (\url{https://www.cosmos.esa.int/gaia}), processed by the \textit{Gaia} Data Processing and Analysis Consortium (DPAC, \url{https://www.cosmos.esa.int/web/gaia/dpac/consortium}). Funding for the DPAC has been provided by national institutions, in particular the institutions participating in the \textit{Gaia} Multilateral Agreement. This research has made use of the SIMBAD database, as well as the VizieR catalogue access tool, CDS, Strasbourg, France. The original description of the VizieR service was published in \citet{ochsbein2000}.

\bibliography{article}

\begin{thebibliography}{}

\bibitem [\protect \citeauthoryear {%
{Anderson}%
\ \BBA {} {Francis}%
}{%
{Anderson}%
\ \BBA {} {Francis}%
}{%
{\protect \APACyear {2012}}%
}]{%
anderson2012}
\APACinsertmetastar {%
anderson2012}%
\begin{APACrefauthors}%
{Anderson}, E.%
\BCBT {}\ \BBA {} {Francis}, C.%
\end{APACrefauthors}%
\unskip\
\newblock
\APACrefYearMonthDay{2012}{{\APACmonth{05}}}{},
\newblock
\unskip
\newblock
\APACjournalVolNumPages{Astron. Lett.}{38}{5}{331-346}.
\PrintBackRefs{\CurrentBib}

\bibitem [\protect \citeauthoryear {%
{Bailer-Jones}%
, {Rybizki}%
, {Fouesneau}%
, {Demleitner}%
\BCBL {}\ \BBA {} {Andrae}%
}{%
{Bailer-Jones}%
\ \protect \BOthers {.}}{%
{\protect \APACyear {2021}}%
}]{%
bailerjones2021}
\APACinsertmetastar {%
bailerjones2021}%
\begin{APACrefauthors}%
{Bailer-Jones}, C\BPBI A\BPBI L.%
, {Rybizki}, J.%
, {Fouesneau}, M.%
, {Demleitner}, M.%
\BCBL {}\ \BBA {} {Andrae}, R.%
\end{APACrefauthors}%
\unskip\
\newblock
\APACrefYearMonthDay{2021}{{\APACmonth{03}}}{},
\newblock
\unskip
\newblock
\APACjournalVolNumPages{AJ}{161}{3}{147}.
\PrintBackRefs{\CurrentBib}

\bibitem [\protect \citeauthoryear {%
{B\"atz}%
, Neuh\"auser%
, Hambaryan%
\BCBL {}\ \BBA {} Hambardzumyan%
}{%
{B\"atz}%
\ \protect \BOthers {.}}{%
{\protect \APACyear {2025}}%
}]{%
baetz2025}
\APACinsertmetastar {%
baetz2025}%
\begin{APACrefauthors}%
{B\"atz}, J.%
, Neuh\"auser, R.%
, Hambaryan, V\BPBI V.%
\BCBL {}\ \BBA {} Hambardzumyan, L\BPBI A.%
\end{APACrefauthors}%
\unskip\
\newblock
\APACrefYearMonthDay{2025}{}{},
\newblock
\unskip
\newblock
\APACjournalVolNumPages{A\&A}{}{}{}.
\PrintBackRefs{\CurrentBib}

\bibitem [\protect \citeauthoryear {%
{B{\'e}dard}%
, {Bergeron}%
, {Brassard}%
\BCBL {}\ \BBA {} {Fontaine}%
}{%
{B{\'e}dard}%
\ \protect \BOthers {.}}{%
{\protect \APACyear {2020}}%
}]{%
bedard2020}
\APACinsertmetastar {%
bedard2020}%
\begin{APACrefauthors}%
{B{\'e}dard}, A.%
, {Bergeron}, P.%
, {Brassard}, P.%
\BCBL {}\ \BBA {} {Fontaine}, G.%
\end{APACrefauthors}%
\unskip\
\newblock
\APACrefYearMonthDay{2020}{{\APACmonth{10}}}{},
\newblock
\unskip
\newblock
\APACjournalVolNumPages{ApJ}{901}{2}{93}.
\PrintBackRefs{\CurrentBib}

\bibitem [\protect \citeauthoryear {%
{Bergeron}%
\ \protect \BOthers {.}}{%
{Bergeron}%
\ \protect \BOthers {.}}{%
{\protect \APACyear {2011}}%
}]{%
bergeron2011}
\APACinsertmetastar {%
bergeron2011}%
\begin{APACrefauthors}%
{Bergeron}, P.%
, {Wesemael}, F.%
, {Dufour}, P.%
\ et al.\end{APACrefauthors}%
\unskip\
\newblock
\APACrefYearMonthDay{2011}{{\APACmonth{08}}}{},
\newblock
\unskip
\newblock
\APACjournalVolNumPages{ApJ}{737}{1}{28}.
\PrintBackRefs{\CurrentBib}

\bibitem [\protect \citeauthoryear {%
{Blaauw}%
}{%
{Blaauw}%
}{%
{\protect \APACyear {1961}}%
}]{%
blaauw1961}
\APACinsertmetastar {%
blaauw1961}%
\begin{APACrefauthors}%
{Blaauw}, A.%
\end{APACrefauthors}%
\unskip\
\newblock
\APACrefYearMonthDay{1961}{{\APACmonth{05}}}{},
\newblock
\unskip
\newblock
\APACjournalVolNumPages{Bulletin Astronomical Institute of the
  Netherlands}{15}{}{265}.
\PrintBackRefs{\CurrentBib}

\bibitem [\protect \citeauthoryear {%
{Blouin}%
, {Dufour}%
\BCBL {}\ \BBA {} {Allard}%
}{%
{Blouin}%
\ \protect \BOthers {.}}{%
{\protect \APACyear {2018}}%
}]{%
blouin2018}
\APACinsertmetastar {%
blouin2018}%
\begin{APACrefauthors}%
{Blouin}, S.%
, {Dufour}, P.%
\BCBL {}\ \BBA {} {Allard}, N\BPBI F.%
\end{APACrefauthors}%
\unskip\
\newblock
\APACrefYearMonthDay{2018}{{\APACmonth{08}}}{},
\newblock
\unskip
\newblock
\APACjournalVolNumPages{ApJ}{863}{2}{184}.
\PrintBackRefs{\CurrentBib}

\bibitem [\protect \citeauthoryear {%
{Bressan}%
\ \protect \BOthers {.}}{%
{Bressan}%
\ \protect \BOthers {.}}{%
{\protect \APACyear {2012}}%
}]{%
bressan2012}
\APACinsertmetastar {%
bressan2012}%
\begin{APACrefauthors}%
{Bressan}, A.%
, {Marigo}, P.%
, {Girardi}, L.%
, {Salasnich}, B.%
, {Dal Cero}, C.%
, {Rubele}, S.%
\BCBL {}\ \BBA {} {Nanni}, A.%
\end{APACrefauthors}%
\unskip\
\newblock
\APACrefYearMonthDay{2012}{{\APACmonth{11}}}{},
\newblock
\unskip
\newblock
\APACjournalVolNumPages{\mnras}{427}{1}{127-145}.
\PrintBackRefs{\CurrentBib}

\bibitem [\protect \citeauthoryear {%
{Carson}%
\ \protect \BOthers {.}}{%
{Carson}%
\ \protect \BOthers {.}}{%
{\protect \APACyear {2013}}%
}]{%
carson2013}
\APACinsertmetastar {%
carson2013}%
\begin{APACrefauthors}%
{Carson}, J.%
, {Thalmann}, C.%
, {Janson}, M.%
\ et al.\end{APACrefauthors}%
\unskip\
\newblock
\APACrefYearMonthDay{2013}{{\APACmonth{02}}}{},
\newblock
\unskip
\newblock
\APACjournalVolNumPages{\apjl}{763}{2}{L32}.
\PrintBackRefs{\CurrentBib}

\bibitem [\protect \citeauthoryear {%
{Castro-Ginard}%
\ \protect \BOthers {.}}{%
{Castro-Ginard}%
\ \protect \BOthers {.}}{%
{\protect \APACyear {2024}}%
}]{%
castro2024}
\APACinsertmetastar {%
castro2024}%
\begin{APACrefauthors}%
{Castro-Ginard}, A.%
, {Penoyre}, Z.%
, {Casey}, A\BPBI R.%
\ et al.\end{APACrefauthors}%
\unskip\
\newblock
\APACrefYearMonthDay{2024}{{\APACmonth{08}}}{},
\newblock
\unskip
\newblock
\APACjournalVolNumPages{\aap}{688}{}{A1}.
\PrintBackRefs{\CurrentBib}

\bibitem [\protect \citeauthoryear {%
{Chen}%
\ \protect \BOthers {.}}{%
{Chen}%
\ \protect \BOthers {.}}{%
{\protect \APACyear {2015}}%
}]{%
chen2015}
\APACinsertmetastar {%
chen2015}%
\begin{APACrefauthors}%
{Chen}, Y.%
, {Bressan}, A.%
, {Girardi}, L.%
, {Marigo}, P.%
, {Kong}, X.%
\BCBL {}\ \BBA {} {Lanza}, A.%
\end{APACrefauthors}%
\unskip\
\newblock
\APACrefYearMonthDay{2015}{{\APACmonth{09}}}{},
\newblock
\unskip
\newblock
\APACjournalVolNumPages{\mnras}{452}{1}{1068-1080}.
\PrintBackRefs{\CurrentBib}

\bibitem [\protect \citeauthoryear {%
{Chen}%
\ \protect \BOthers {.}}{%
{Chen}%
\ \protect \BOthers {.}}{%
{\protect \APACyear {2014}}%
}]{%
chen2014}
\APACinsertmetastar {%
chen2014}%
\begin{APACrefauthors}%
{Chen}, Y.%
, {Girardi}, L.%
, {Bressan}, A.%
, {Marigo}, P.%
, {Barbieri}, M.%
\BCBL {}\ \BBA {} {Kong}, X.%
\end{APACrefauthors}%
\unskip\
\newblock
\APACrefYearMonthDay{2014}{{\APACmonth{11}}}{},
\newblock
\unskip
\newblock
\APACjournalVolNumPages{\mnras}{444}{3}{2525-2543}.
\PrintBackRefs{\CurrentBib}

\bibitem [\protect \citeauthoryear {%
{Drummond}%
}{%
{Drummond}%
}{%
{\protect \APACyear {2014}}%
}]{%
drummond2014}
\APACinsertmetastar {%
drummond2014}%
\begin{APACrefauthors}%
{Drummond}, J\BPBI D.%
\end{APACrefauthors}%
\unskip\
\newblock
\APACrefYearMonthDay{2014}{{\APACmonth{03}}}{},
\newblock
\unskip
\newblock
\APACjournalVolNumPages{\aj}{147}{3}{65}.
\PrintBackRefs{\CurrentBib}

\bibitem [\protect \citeauthoryear {%
{Gaia Collaboration}%
\ \protect \BOthers {.}}{%
{Gaia Collaboration}%
\ \protect \BOthers {.}}{%
{\protect \APACyear {2023}}%
}]{%
gaiadr3}
\APACinsertmetastar {%
gaiadr3}%
\begin{APACrefauthors}%
{Gaia Collaboration}%
, {Vallenari}, A.%
, {Brown}, A\BPBI G\BPBI A.%
\ et al.\end{APACrefauthors}%
\unskip\
\newblock
\APACrefYearMonthDay{2023}{{\APACmonth{06}}}{},
\newblock
\unskip
\newblock
\APACjournalVolNumPages{A\&A}{674}{}{A1}.
\PrintBackRefs{\CurrentBib}

\bibitem [\protect \citeauthoryear {%
{Hartkopf}%
, {Mason}%
\BCBL {}\ \BBA {} {Worley}%
}{%
{Hartkopf}%
\ \protect \BOthers {.}}{%
{\protect \APACyear {2001}}%
}]{%
hartkopf2001}
\APACinsertmetastar {%
hartkopf2001}%
\begin{APACrefauthors}%
{Hartkopf}, W\BPBI I.%
, {Mason}, B\BPBI D.%
\BCBL {}\ \BBA {} {Worley}, C\BPBI E.%
\end{APACrefauthors}%
\unskip\
\newblock
\APACrefYearMonthDay{2001}{{\APACmonth{12}}}{},
\newblock
\unskip
\newblock
\APACjournalVolNumPages{\aj}{122}{6}{3472-3479}.
\PrintBackRefs{\CurrentBib}

\bibitem [\protect \citeauthoryear {%
{Holberg}%
\ \BBA {} {Bergeron}%
}{%
{Holberg}%
\ \BBA {} {Bergeron}%
}{%
{\protect \APACyear {2006}}%
}]{%
holberg2006}
\APACinsertmetastar {%
holberg2006}%
\begin{APACrefauthors}%
{Holberg}, J\BPBI B.%
\BCBT {}\ \BBA {} {Bergeron}, P.%
\end{APACrefauthors}%
\unskip\
\newblock
\APACrefYearMonthDay{2006}{{\APACmonth{09}}}{},
\newblock
\unskip
\newblock
\APACjournalVolNumPages{AJ}{132}{}{1221}.
\PrintBackRefs{\CurrentBib}

\bibitem [\protect \citeauthoryear {%
{Johnson}%
}{%
{Johnson}%
}{%
{\protect \APACyear {2004}}%
}]{%
johnson2004}
\APACinsertmetastar {%
johnson2004}%
\begin{APACrefauthors}%
{Johnson}, D\BPBI O.%
\end{APACrefauthors}%
\unskip\
\newblock
\APACrefYearMonthDay{2004}{{\APACmonth{12}}}{},
\newblock
\unskip
\newblock
\APACjournalVolNumPages{Journal of Astronomical Data}{10}{}{3}.
\PrintBackRefs{\CurrentBib}

\bibitem [\protect \citeauthoryear {%
{Kowalski}%
\ \BBA {} {Saumon}%
}{%
{Kowalski}%
\ \BBA {} {Saumon}%
}{%
{\protect \APACyear {2006}}%
}]{%
kowalski2006}
\APACinsertmetastar {%
kowalski2006}%
\begin{APACrefauthors}%
{Kowalski}, P\BPBI M.%
\BCBT {}\ \BBA {} {Saumon}, D.%
\end{APACrefauthors}%
\unskip\
\newblock
\APACrefYearMonthDay{2006}{{\APACmonth{11}}}{},
\newblock
\unskip
\newblock
\APACjournalVolNumPages{ApJL}{651}{}{L137}.
\PrintBackRefs{\CurrentBib}

\bibitem [\protect \citeauthoryear {%
{Marigo}%
\ \protect \BOthers {.}}{%
{Marigo}%
\ \protect \BOthers {.}}{%
{\protect \APACyear {2017}}%
}]{%
marigo2017}
\APACinsertmetastar {%
marigo2017}%
\begin{APACrefauthors}%
{Marigo}, P.%
, {Girardi}, L.%
, {Bressan}, A.%
\ et al.\end{APACrefauthors}%
\unskip\
\newblock
\APACrefYearMonthDay{2017}{{\APACmonth{01}}}{},
\newblock
\unskip
\newblock
\APACjournalVolNumPages{\apj}{835}{1}{77}.
\PrintBackRefs{\CurrentBib}

\bibitem [\protect \citeauthoryear {%
{Mason}%
, {Wycoff}%
, {Hartkopf}%
, {Douglass}%
\BCBL {}\ \BBA {} {Worley}%
}{%
{Mason}%
\ \protect \BOthers {.}}{%
{\protect \APACyear {2001}}%
}]{%
mason2001}
\APACinsertmetastar {%
mason2001}%
\begin{APACrefauthors}%
{Mason}, B\BPBI D.%
, {Wycoff}, G\BPBI L.%
, {Hartkopf}, W\BPBI I.%
, {Douglass}, G\BPBI G.%
\BCBL {}\ \BBA {} {Worley}, C\BPBI E.%
\end{APACrefauthors}%
\unskip\
\newblock
\APACrefYearMonthDay{2001}{{\APACmonth{12}}}{},
\newblock
\unskip
\newblock
\APACjournalVolNumPages{AJ}{122}{6}{3466-3471}.
\PrintBackRefs{\CurrentBib}

\bibitem [\protect \citeauthoryear {%
{Mugrauer}%
}{%
{Mugrauer}%
}{%
{\protect \APACyear {2019}}%
}]{%
mugrauer2019}
\APACinsertmetastar {%
mugrauer2019}%
\begin{APACrefauthors}%
{Mugrauer}, M.%
\end{APACrefauthors}%
\unskip\
\newblock
\APACrefYearMonthDay{2019}{{\APACmonth{12}}}{},
\newblock
\unskip
\newblock
\APACjournalVolNumPages{\mnras}{490}{4}{5088-5102}.
\PrintBackRefs{\CurrentBib}

\bibitem [\protect \citeauthoryear {%
{Mugrauer}%
, {Avila}%
\BCBL {}\ \BBA {} {Guirao}%
}{%
{Mugrauer}%
\ \protect \BOthers {.}}{%
{\protect \APACyear {2014}}%
}]{%
mugrauer2014}
\APACinsertmetastar {%
mugrauer2014}%
\begin{APACrefauthors}%
{Mugrauer}, M.%
, {Avila}, G.%
\BCBL {}\ \BBA {} {Guirao}, C.%
\end{APACrefauthors}%
\unskip\
\newblock
\APACrefYearMonthDay{2014}{{\APACmonth{01}}}{},
\newblock
\unskip
\newblock
\APACjournalVolNumPages{AN}{335}{4}{417}.
\PrintBackRefs{\CurrentBib}

\bibitem [\protect \citeauthoryear {%
{Mugrauer}%
, {Kollak}%
, {Pietsch}%
\BCBL {}\ \BBA {} {Michel}%
}{%
{Mugrauer}%
\ \protect \BOthers {.}}{%
{\protect \APACyear {2025}}%
}]{%
mugrauer2025}
\APACinsertmetastar {%
mugrauer2025}%
\begin{APACrefauthors}%
{Mugrauer}, M.%
, {Kollak}, A\BHBI K.%
, {Pietsch}, L.%
\BCBL {}\ \BBA {} {Michel}, K\BHBI U.%
\end{APACrefauthors}%
\unskip\
\newblock
\APACrefYearMonthDay{2025}{{\APACmonth{06}}}{},
\newblock
\unskip
\newblock
\APACjournalVolNumPages{Astronomische Nachrichten}{346}{5}{e70005}.
\PrintBackRefs{\CurrentBib}

\bibitem [\protect \citeauthoryear {%
{Ochsenbein}%
, {Bauer}%
\BCBL {}\ \BBA {} {Marcout}%
}{%
{Ochsenbein}%
\ \protect \BOthers {.}}{%
{\protect \APACyear {2000}}%
}]{%
ochsbein2000}
\APACinsertmetastar {%
ochsbein2000}%
\begin{APACrefauthors}%
{Ochsenbein}, F.%
, {Bauer}, P.%
\BCBL {}\ \BBA {} {Marcout}, J.%
\end{APACrefauthors}%
\unskip\
\newblock
\APACrefYearMonthDay{2000}{{\APACmonth{04}}}{},
\newblock
\unskip
\newblock
\APACjournalVolNumPages{A\&A,S}{143}{}{23-32}.
\PrintBackRefs{\CurrentBib}

\bibitem [\protect \citeauthoryear {%
{Pastorelli}%
\ \protect \BOthers {.}}{%
{Pastorelli}%
\ \protect \BOthers {.}}{%
{\protect \APACyear {2020}}%
}]{%
pastorelli2020}
\APACinsertmetastar {%
pastorelli2020}%
\begin{APACrefauthors}%
{Pastorelli}, G.%
, {Marigo}, P.%
, {Girardi}, L.%
\ et al.\end{APACrefauthors}%
\unskip\
\newblock
\APACrefYearMonthDay{2020}{{\APACmonth{11}}}{},
\newblock
\unskip
\newblock
\APACjournalVolNumPages{\mnras}{498}{3}{3283-3301}.
\PrintBackRefs{\CurrentBib}

\bibitem [\protect \citeauthoryear {%
{Pastorelli}%
\ \protect \BOthers {.}}{%
{Pastorelli}%
\ \protect \BOthers {.}}{%
{\protect \APACyear {2019}}%
}]{%
pastorelli2019}
\APACinsertmetastar {%
pastorelli2019}%
\begin{APACrefauthors}%
{Pastorelli}, G.%
, {Marigo}, P.%
, {Girardi}, L.%
\ et al.\end{APACrefauthors}%
\unskip\
\newblock
\APACrefYearMonthDay{2019}{{\APACmonth{06}}}{},
\newblock
\unskip
\newblock
\APACjournalVolNumPages{\mnras}{485}{4}{5666-5692}.
\PrintBackRefs{\CurrentBib}

\bibitem [\protect \citeauthoryear {%
{Pourbaix}%
\ \protect \BOthers {.}}{%
{Pourbaix}%
\ \protect \BOthers {.}}{%
{\protect \APACyear {2004}}%
}]{%
pourbaix2004}
\APACinsertmetastar {%
pourbaix2004}%
\begin{APACrefauthors}%
{Pourbaix}, D.%
, {Tokovinin}, A\BPBI A.%
, {Batten}, A\BPBI H.%
\ et al.\end{APACrefauthors}%
\unskip\
\newblock
\APACrefYearMonthDay{2004}{{\APACmonth{09}}}{},
\newblock
\unskip
\newblock
\APACjournalVolNumPages{\aap}{424}{}{727-732}.
\PrintBackRefs{\CurrentBib}

\bibitem [\protect \citeauthoryear {%
{Tang}%
\ \protect \BOthers {.}}{%
{Tang}%
\ \protect \BOthers {.}}{%
{\protect \APACyear {2014}}%
}]{%
tang2014}
\APACinsertmetastar {%
tang2014}%
\begin{APACrefauthors}%
{Tang}, J.%
, {Bressan}, A.%
, {Rosenfield}, P.%
, {Slemer}, A.%
, {Marigo}, P.%
, {Girardi}, L.%
\BCBL {}\ \BBA {} {Bianchi}, L.%
\end{APACrefauthors}%
\unskip\
\newblock
\APACrefYearMonthDay{2014}{{\APACmonth{12}}}{},
\newblock
\unskip
\newblock
\APACjournalVolNumPages{\mnras}{445}{4}{4287-4305}.
\PrintBackRefs{\CurrentBib}

\bibitem [\protect \citeauthoryear {%
{Tody}%
}{%
{Tody}%
}{%
{\protect \APACyear {1993}}%
}]{%
tody1993}
\APACinsertmetastar {%
tody1993}%
\begin{APACrefauthors}%
{Tody}, D.%
\end{APACrefauthors}%
\unskip\
\newblock
\APACrefYearMonthDay{1993}{{\APACmonth{01}}}{},
\newblock
{\BBOQ}\APACrefatitle {{IRAF in the Nineties}} {{IRAF in the Nineties}}.{\BBCQ}
\newblock
\BIn{} R\BPBI J.~{Hanisch}, R\BPBI J\BPBI V.~{Brissenden}\BCBL {}\ \BBA {}
  J.~{Barnes}\ (\BEDS), \APACrefbtitle {Astronomical Data Analysis Software and
  Systems II} {Astronomical Data Analysis Software and Systems II}\ \BVOL~52,
  \BPG~173.
\PrintBackRefs{\CurrentBib}

\bibitem [\protect \citeauthoryear {%
{Tremblay}%
, {Bergeron}%
\BCBL {}\ \BBA {} {Gianninas}%
}{%
{Tremblay}%
\ \protect \BOthers {.}}{%
{\protect \APACyear {2011}}%
}]{%
tremblay2011}
\APACinsertmetastar {%
tremblay2011}%
\begin{APACrefauthors}%
{Tremblay}, P\BHBI E.%
, {Bergeron}, P.%
\BCBL {}\ \BBA {} {Gianninas}, A.%
\end{APACrefauthors}%
\unskip\
\newblock
\APACrefYearMonthDay{2011}{{\APACmonth{04}}}{},
\newblock
\unskip
\newblock
\APACjournalVolNumPages{ApJ}{730}{}{128}.
\PrintBackRefs{\CurrentBib}

\bibitem [\protect \citeauthoryear {%
{Wenger}%
\ \protect \BOthers {.}}{%
{Wenger}%
\ \protect \BOthers {.}}{%
{\protect \APACyear {2000}}%
}]{%
wenger2000}
\APACinsertmetastar {%
wenger2000}%
\begin{APACrefauthors}%
{Wenger}, M.%
, {Ochsenbein}, F.%
, {Egret}, D.%
\ et al.\end{APACrefauthors}%
\unskip\
\newblock
\APACrefYearMonthDay{2000}{{\APACmonth{04}}}{},
\newblock
\unskip
\newblock
\APACjournalVolNumPages{A\&A,S}{143}{}{9-22}.
\PrintBackRefs{\CurrentBib}

\end{thebibliography}

\end{document}